\documentclass[a4paper,12pt]{article}
\usepackage[utf8]{inputenc}
\usepackage{multicol} 
\usepackage[toc,page]{appendix}
\usepackage{graphicx}
\usepackage{caption}
\usepackage{subcaption}
\usepackage{amsmath,amssymb}
\usepackage[margin=0.80in]{geometry}
\usepackage{color}
\usepackage[colorlinks = true,
            linkcolor = blue,
            urlcolor  = blue,
            citecolor = blue,
            anchorcolor = blue]{hyperref}
\usepackage{float}
\usepackage[usenames,dvipsnames]{xcolor}
\usepackage{gensymb}
\usepackage{cite}
\usepackage{authblk}

\newcommand{\specialcell}[2][c]{\begin{tabular}[#1]{@{}c@{}}#2\end{tabular}}
\DeclareMathOperator\erf{Erfc}

\title{Experimental calibration of a Cahn-Hilliard phase-field model for phase
transformations in Li-Sn electrodes}
\author[1]{Srivatsan Hulikal\thanks{srivatsan\_hulikal@brown.edu}}
\author[1]{Chun-Hao Chen \thanks{chun-hao\_chen@brown.edu}}
\author[1]{Eric Chason \thanks{eric\_chason\_phd@brown.edu@brown.edu}}
\author[1]{Allan Bower \thanks{Allan\_Bower@brown.edu @brown.edu}}
\affil[1]{School of Engineering, Brown University, Providence, RI 02912}
\date{}

\begin{document}
\maketitle
\begin{abstract}
Experiments, numerical simulations, and analytical calculations are used to
calibrate a diffuse-interface Cahn-Hilliard model of Li-Sn thin film electrodes
that experience a transformation from Sn to Li$_2$Sn$_5$. It is shown that a
concentration-gradient dependent mobility can be used in the Cahn-Hilliard
equation to give the interface a finite mobility and capture its nonequilibrium
behavior. Comparing experiments and simulations, the free-energy of Li-Sn,
diffusivity of Li in Sn and Li$_2$Sn$_5$, the exchange current density for the
surface reaction, and the mobility of the Sn/Li$_2$Sn$_5$ interface are
extracted. The implications of finite interface mobility for practical battery
electrodes are discussed.
\end{abstract}

\iftrue
\section{Introduction}

Diffusional phase transformations play an important role in many Li-ion battery
electrodes. Cathode materials such as LiFePO$_4$, Li$_x$CoO$_2$, and
Li$_x$MnO$_4$ \cite{RN31}, and anodes such as graphite, silicon
\cite{RN23,RN118}, and tin (Sn) \cite{RN155} undergo phase transformations upon
lithiation and delithiation. The charge-discharge dynamics of a battery depends
critically on the nucleation and propagation of the phases. In some materials,
stresses developed due to volume change during phase transformation and
mismatch strain across the phase boundary are significant and affect transport
properties \cite{RN41,RN166}. The stresses can also lead to the degradation and
failure of the electrode thus affecting the life of the battery
\cite{RN85,RN167}. Interface propagation is accompanied by energy dissipation
which limits the efficiency \cite{RN116}. Thus, a good understanding of these
phase transformations using experiments and modeling is imperative in designing
better batteries.

Phase-field models have been used widely in studying phase transformation
phenomena \cite{RN157,RN158,RN14,RN25,RN144,RN17,RN48}. In these models
\cite{RN157,RN158}, a continuous field variable represents the phase of the
material: for example, in the Cahn-Hilliard model \cite{RN154}, the
concentration of the diffusing species is the phase variable. The free-energy
is assumed to be a non-convex function of the concentration, so that the system
naturally phase-separates.  A contribution to the free-energy from the
concentration gradient results in the formation of a diffuse phase-boundary
between phases. Phase-field models have several advantages. Since the phase
variable is continuous, explicit interface tracking is unnecessary. Their
variational nature aids the development of stable numerical methods. In
addition, Cahn-Hilliard models can also nucleate new phases.

Cahn-Hilliard was first used to study battery materials by Han et al.
\cite{RN14} who determined the diffusivity of Li in Olivine LiFePO$_4$ using
Galvanostatic Intermittent Titration Technique (GITT) and Potentiostatic
Intermittent Titration Technique (PITT) experiments and evaluated the effect of
gradient energy on transport.  Since then, it has been used widely to study
several aspects of phase separation phenomena.  Using the Cahn-Hilliard model
with new insertion kinetics for the surface-reaction, Singh et al. \cite{RN17}
studied the effect of anisotropic mobility on the intercalation dynamics and
found a surface-reaction-limited regime where phase-boundaries move along the
electrode surface as opposed to the shirking-core models where they move within
the bulk.  Burch/Bazant \cite{RN144} showed that the spinodal and miscibility
gaps become smaller with decreasing particle size leading to phase-separation
suppression in nanoparticles. Bai et al. \cite{RN140} used an averaged
phase-field model and showed the disappearance of spinodal resulting in
homogeneous lithiation of nanoscale LiFePO$_4$ particles which may explain their
high-rate behaviour.  The effect of stress has also been included, for example
Cogswell/Bazant \cite{RN25} studied how coherency strain influences
phase-boundary orientation and microstructure morphology in LiFePO$_4$.
Coupling a phase-field model with elasticity, Tang et al. \cite{RN15} looked at
the effect of particle size, misfit strain, and electric overpotential in the
amorphization of nanoscale olivines.  Based on a ``microforce balance"
approach, Anand \cite{RN48} extended the Cahn-Hilliard model to account for
large elastic-plastic deformations.  Using this, Di Leo et al.  \cite{RN169}
studied the effect of plasticity on state-of-charge and energy dissipation in
amorphous Si electrodes. Apart from single electrode particles, phase-field
models have also been used to study porous electrodes \cite{RN127}.

Cahn-Hilliard models contain a number of material parameters that must be
determined experimentally. In this paper, our goal is to calibrate a
Cahn-Hilliard model of Li-Sn by careful comparison of predictions with
experiments.  We use the Li-Sn system as a representative example of a
practical battery electrode material that experiences several phase
transformations during Li insertion and removal. Sn, with a large theoretical
gravimetric capacity of about 990 mAh g$^{-1}$ is a promising anode material
for Li-ion batteries.  Li-Sn serves as a canonical system because upon
lithiation, it undergoes a series of crystalline-crystalline phase
transformations that are reversible \cite{RN155,RN170,RN171,RN172,RN173}.
Wen/Huggins \cite{RN170} showed that the Li-Sn system has six phases at high
temperature: LiSn, Li$_7$Sn$_3$, Li$_5$Sn$_2$, Li$_{13}$Sn$_5$, Li$_7$Sn$_2$,
and Li$_{22}$Sn$_5$.  Using in-situ XRD, Rhodes et al. \cite{RN155} identified
phases Li$_2$Sn$_5$, LiSn, and Li$_{22}$Sn$_5$ during lithiation cycles at room
temperature. Many studies have looked at the equilibrium properties of Li-Sn
and phase diagrams have been constructed (see \cite{RN171,RN172,RN173} and
references therein). For example, plateau potentials of the successive phase
transformations at 25 \degree C range from 0.76 V to 0.38 V (against Li/Li$^+$)
\cite{RN173}.  Apart from equilibrium properties, kinetic parameters have also
been measured.  The chemical diffusivity of Li in the various phases at 415
\degree C is of the order of 10$^{-5}$ cm$^2$sec$^{-1}$ \cite{RN119}. At room
temperature, diffusivities of 10$^{-8}-10^{-7}$ cm$^2$sec$^{-1}$ for
Li$_{0.7}$Sn and Li$_{2.33}$Sn have been reported \cite{RN174}. Much smaller
diffusivities (10$^{-16}-10^{-14}$ cm$^2$sec$^{-1}$) have been observed for Sn
\cite{RN175}.

Here, we focus on transformations between Sn and Li$_2$Sn$_5$ which are the
first two phases to form at room temperature. We measure the in-situ
variation of current and voltage in Sn thin film electrodes with Li
counter-electrode subjected to lithiation and delithiation (Section
\ref{sec:experiments}). Comparing predictions of a Cahn-Hilliard model with
experimental observations, we determine the free-energy of Li-Sn as a function
of Li concentration, the diffusivity of Li in the first two phases (Sn and
Li$_2$Sn$_5$), the surface-reaction parameters in the Butler-Volmer equations,
and most importantly, the interface mobility (Section \ref{sec:calibration}).
As far as we know, these are the first measurements of diffusivity in
Li$_2$Sn$_5$, the surface-reaction rate for the insertion reaction at the
electrode/electrolyte interface, and mobility of the Sn-Li$_2$Sn$_5$ interface.
The standard Cahn-Hilliard equations predict that the interface has infinite
mobility thus ignoring the nonequilibrium interface behaviour which can be
important in nanoscale electrodes. Following Langer and Sekerka \cite{RN153},
we show that the Cahn-Hilliard equations can be modified to model
interface-limited processes by including a concentration-gradient dependent
mobility of Li (Section \ref{sec:cahnHilliard}). We show representative
examples that elucidate the general behavior of the modified Cahn-Hilliard
model (Section \ref{sec:modelBehavior}).

To better understand interface behavior, we find it helpful to study the
sharp-interface limit of the Cahn-Hilliard equations.  Accordingly, we present
results of a perturbation analysis of the modified Cahn-Hilliard equations
based on a paper by Langer and Sekerka \cite{RN153}.  We derive a general
energy-based sharp-interface model that identifies the fluxes and conjugate
forces for dissipative processes in the bulk and at the interface. A comparison
with the perturbation analysis results reveals that the kinetic relations for
the interface implied by the Cahn-Hilliard equations are a particular special
case of a more general class (Section \ref{sec:sharpInterface}). We discuss the
implications of interface mobility for the charge/discharge dynamics and energy
efficiency of a battery (Section \ref{sec:discussion}).

\fi

\iftrue
\section{Experiment}\label{sec:experiments}

We briefly review the experimental procedure used in this study. The apparatus
consists of a Li-ion half-cell with a thin-film Sn working electrode and a Li
metal foil counter-electrode as illustrated in Figure
\ref{fig:electrodeSystemSchematic}. The Sn thin film was deposited on
silica quartz wafers ($50.8$ mm diameter, $450-500 \mu$m thick, double-sided
polished). Prior to film deposition, the silica wafers were cleaned with
Acetone, Methonal, Isopropenal and de-ionized (DI) water for 5 minutes each in
sequence, followed by drying with compressed nitrogen gas. A 25 nm adhesion Ti
layer and 50 nm current-collected Cu layer were deposited on one side of the
wafers via physical vapor deposition (PVD) at working pressure below
$2\times10^{-6}$ Torr. Then, a Sn anode layer with thickness between 200 nm and
2$\mu$m was electroplated on the Cu layers with a commercial Sn electroplating
solution (Solderon SC, produced by Rohm \& Haas). Before electroplating, the
wafers were etched by 98 \% sulfuric acid to remove the native oxide layer on
the surface. After the Sn film growth, the samples were cleaned with acetone
and DI water in sequence for 5 minutes, and then dried with compressed nitrogen
gas. Post sample fabrication, the samples were stored in an Ar-filled glove
box, and both moisture and oxygen were below 0.1ppm. A 0.5 mm thick Li foil was
used as both counter and reference electrode. The separator was Celgard C480
(Celgard Inc., Charlotte, NC). The electrolyte composition was 1.2 M LiPF6 in
EC/DEC with ratio 1:2 (wt.  \%) (BASF Corp. A6 Series). The electrochemical
measurements were done using Multistat 1470E (Solartron Analytical) at room
temperature. 

\begin{figure}[H]
\centering
\includegraphics[scale=0.5]{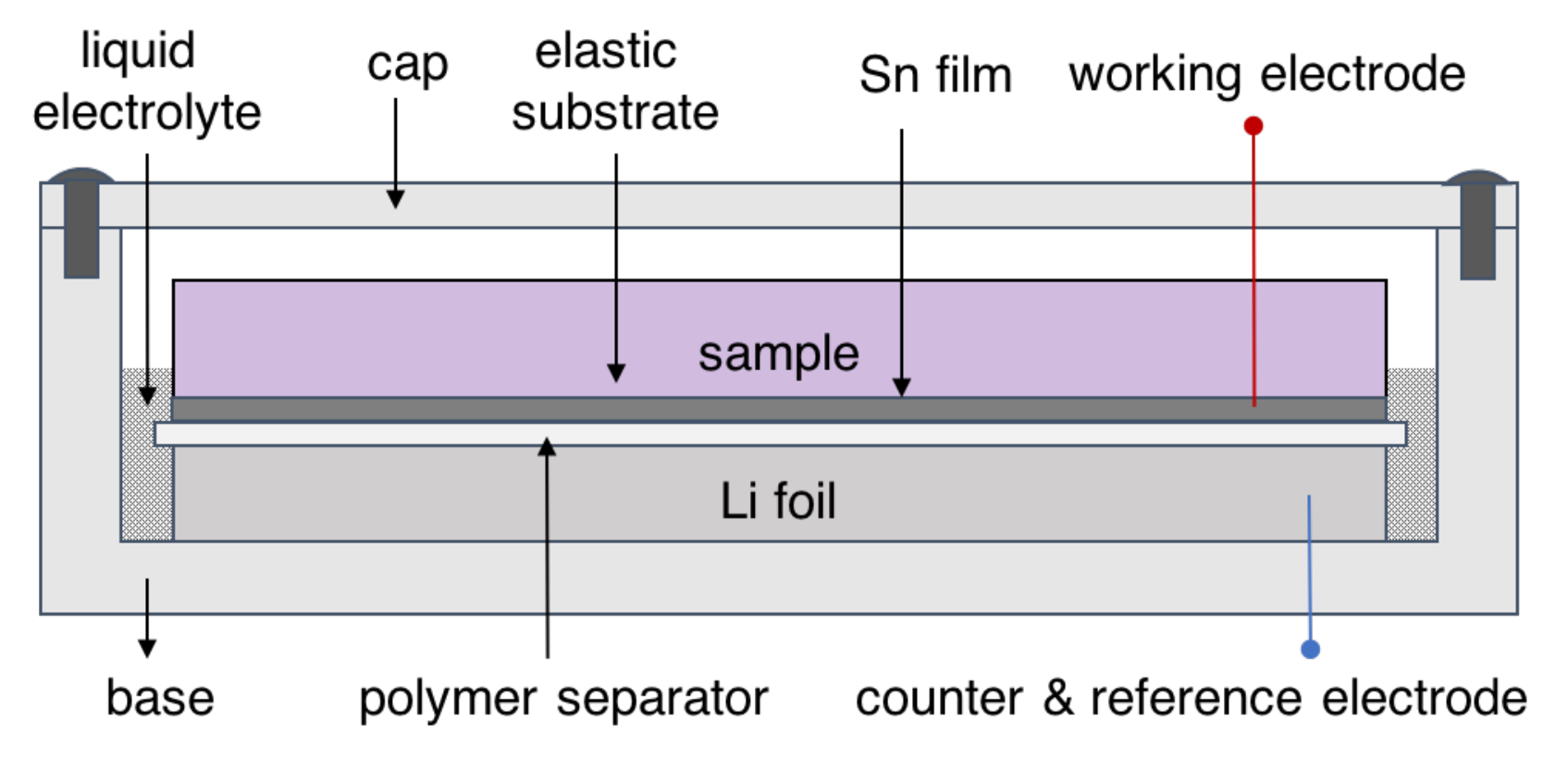}
\caption{A schematic representation of the Li-Sn thin film system used in our
experiments. The apparatus consists of a Li-ion half-cell with a thin-film Sn
working electrode on silica quartz wafer substrate and a Li metal foil
counter-electrode.}
\label{fig:electrodeSystemSchematic}
\end{figure}

Prior to all experiments, the Sn electrode was first lithiated
galvanostatically with C/20 rate from open-circuit potential at 2.75 V to 0.8 V
and then the potential was held at 0.8 V for 20 hours to form a stable SEI
layer to minimize the influence of the side reactions.  We then conducted three
kinds of experiments, PITT \cite{RN177}, Potentiostatic
lithiation/delithiation, and Galvanostatic lithiation/delithiation. 

PITT was done to study the Sn phase. The potential was decreased from 0.8 V
(after SEI growth phase) in steps of 0.02 V till 0.7 V at which the
Li$_2$Sn$_5$ phase nucleated (Figure
\ref{fig:timeVoltageCurrentExperimentalPlotFirstLithiation}). In each step, the
potential applied was held until the current fell below 0.05 mA g$^{-1}$ (about
C/1800). The transient current evolution following the voltage steps is used to
determine the diffusivity and the Butler-Volmer parameter as discussed in
detail in Section \ref{sec:calibration}.  In addition, the free-energy of Li-Sn
as a function of Li concentration was determined from the steady-state
voltage-charge measurements.

To determine the exchange current density of Li$_2$Sn$_5$, the potential was
lowered to 0.65 V to nucleate the Li$_2$Sn$_5$ phase and grow it to approximately
half the film thickness. The film was then allowed to reach equilibrium.
Subsequently, a step voltage change was applied (from the equilibrium voltage)
and the transient current was used to determine the Butler-Volmer parameter.

Galvanostatic lithiation/delithiation was used to determine the
interface mobility. First, the potential was stepped down from 0.8 V to 0.665
V, sufficiently low to nucleate and grow the Li$_2$Sn$_5$ phase. Then, the film
was lithiated/delithiated at currents from C/2500 to C/625 allowing to it
equilibriate in between. The plateau potential as a function of the current was
used to determine the interface mobility.

Figure \ref{fig:experimentalResults} shows representative current and
voltage evolution with time in PITT and Potentiostatic experiments and Table
\ref{table:experiments} lists details of the various experiments used in the
calibration.

\begin{figure}[h]
\centering
\begin{subfigure}[t]{0.49\textwidth}
\includegraphics[width=\textwidth]{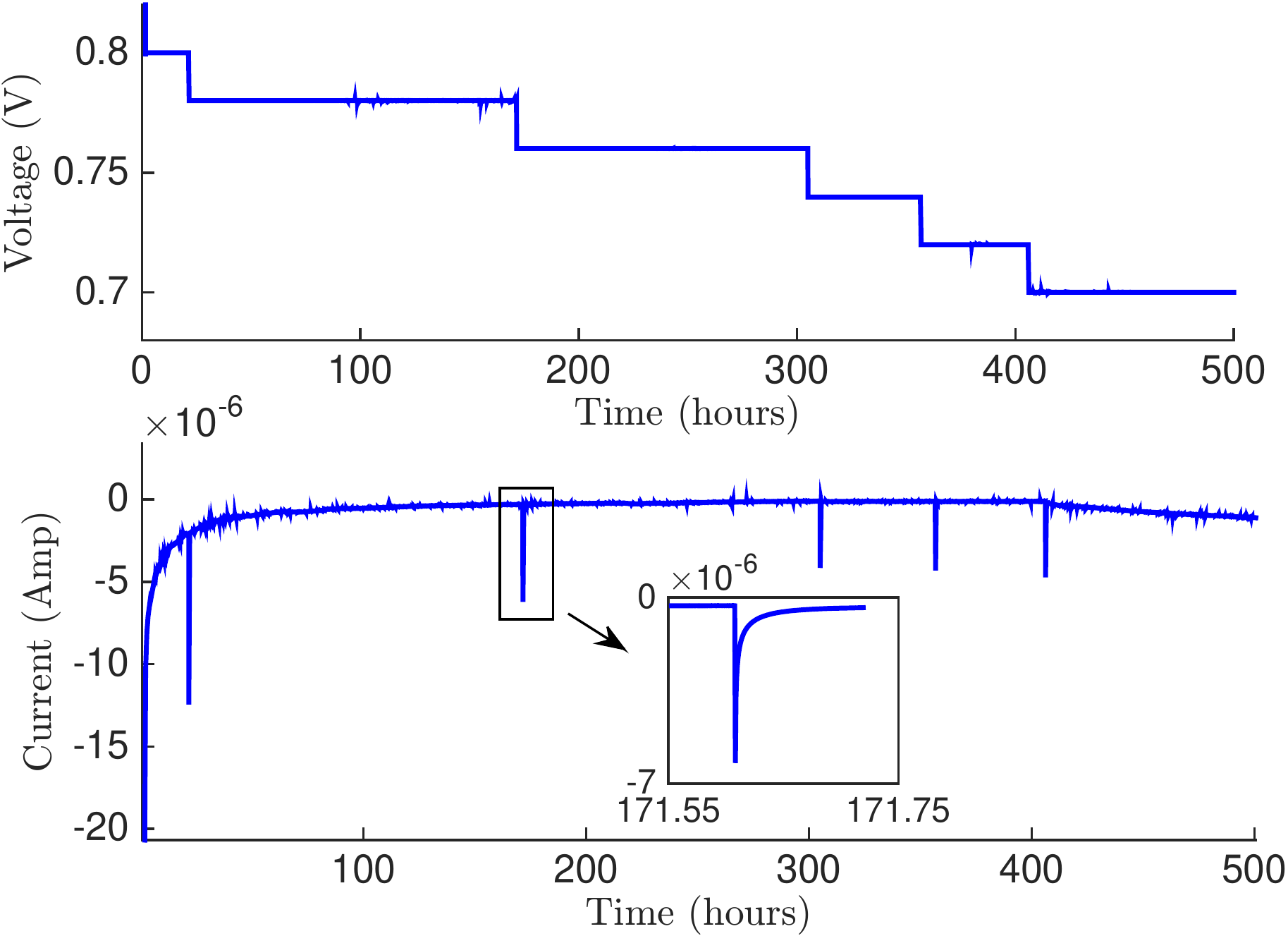}
\caption{}
\label{fig:timeVoltageCurrentExperimentalPlotFirstLithiation}
\end{subfigure} 
\begin{subfigure}[t]{0.49\textwidth}
\includegraphics[width=\textwidth]{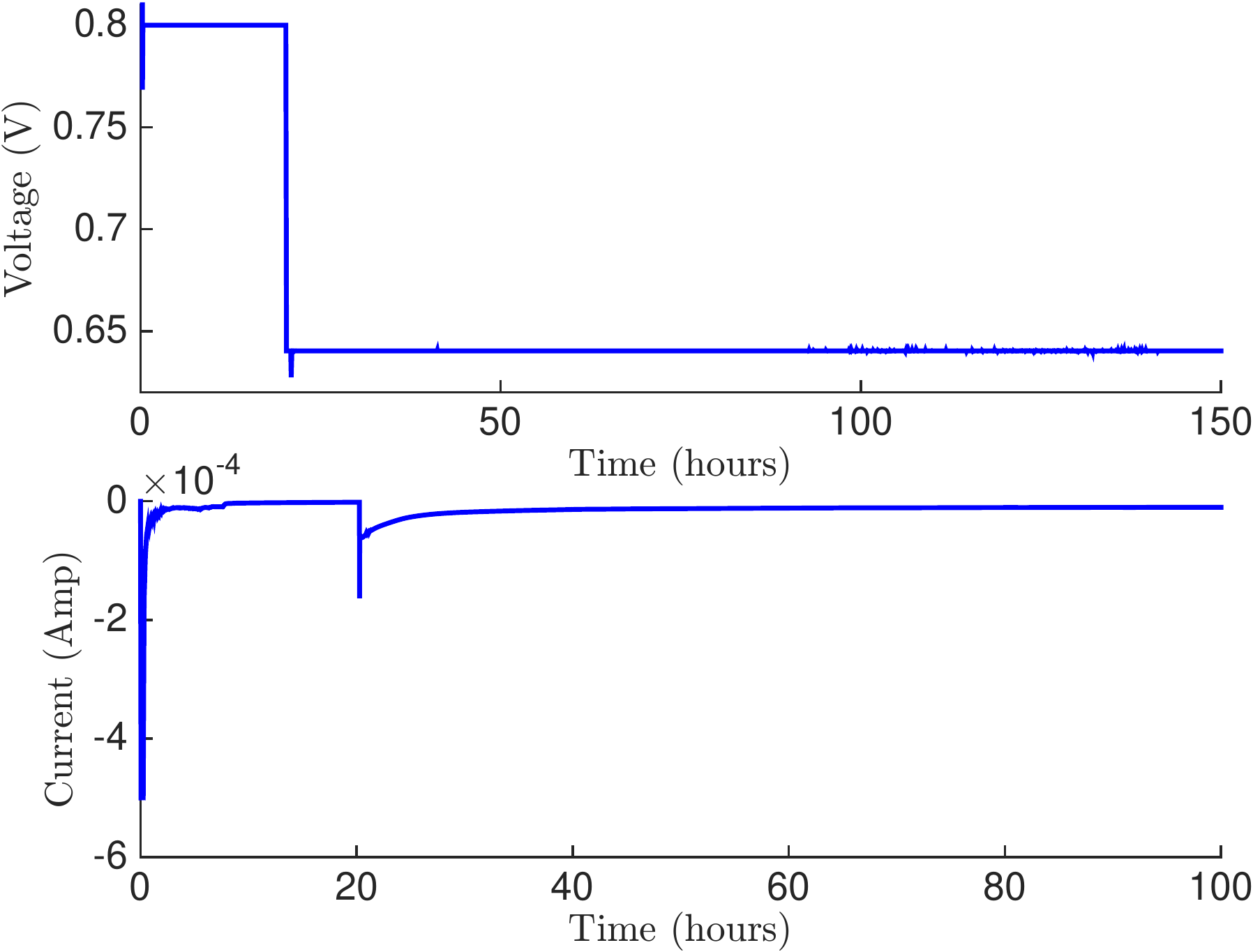}
\caption{}
\label{fig:timeVoltageCurrentExperimentalPlotCell103}
\end{subfigure} 
\caption{Voltage and current evolution in (a) PITT and (b)
Potentiostatic experiments. These and other similar experiments (see Table
\ref{table:experiments} and Section \ref{sec:calibration}) are used to
determine the parameters in the Cahn-Hilliard model and compare its
predictions.}
\label{fig:experimentalResults}
\end{figure}

\begin{table}[H]
\centering
\begin{tabular}{|c|c|c|c|}
\hline
\textbf{Experiment}  & \textbf{Parmeters calibrated} & \textbf{Film
thickness} & \textbf{Range}  \\ \hline
PITT & \specialcell{$D,M_0,i_0$ of Sn \\ Free-energy $G_0(c)$} & 187 nm  & 0.8V to 0.7V - steps of 0.02V \\ \hline 
Potentiostatic &  $i_0$ of Li$_2$Sn$_5$ & $ 1.85 \mu$m & 0.7V, 0.8V \\ \hline 
Galvanostatic   & $i_0,\chi$ & $1.85 \mu$m & C/2500 to C/625  \\ \hline
\end{tabular}
\caption{Experiments used to determine the parameters in the Cahn-Hilliard
model. See Sections \ref{sec:cahnHilliard} and \ref{sec:calibration} for more
details.}
\label{table:experiments}
\end{table}

\fi 

\iftrue
\section{Phase field model of Li-Sn thin film electrodes}
\label{sec:cahnHilliard}

\begin{figure}[H]
\centering
\begin{subfigure}[c]{0.49\textwidth}
\includegraphics[width=\textwidth]{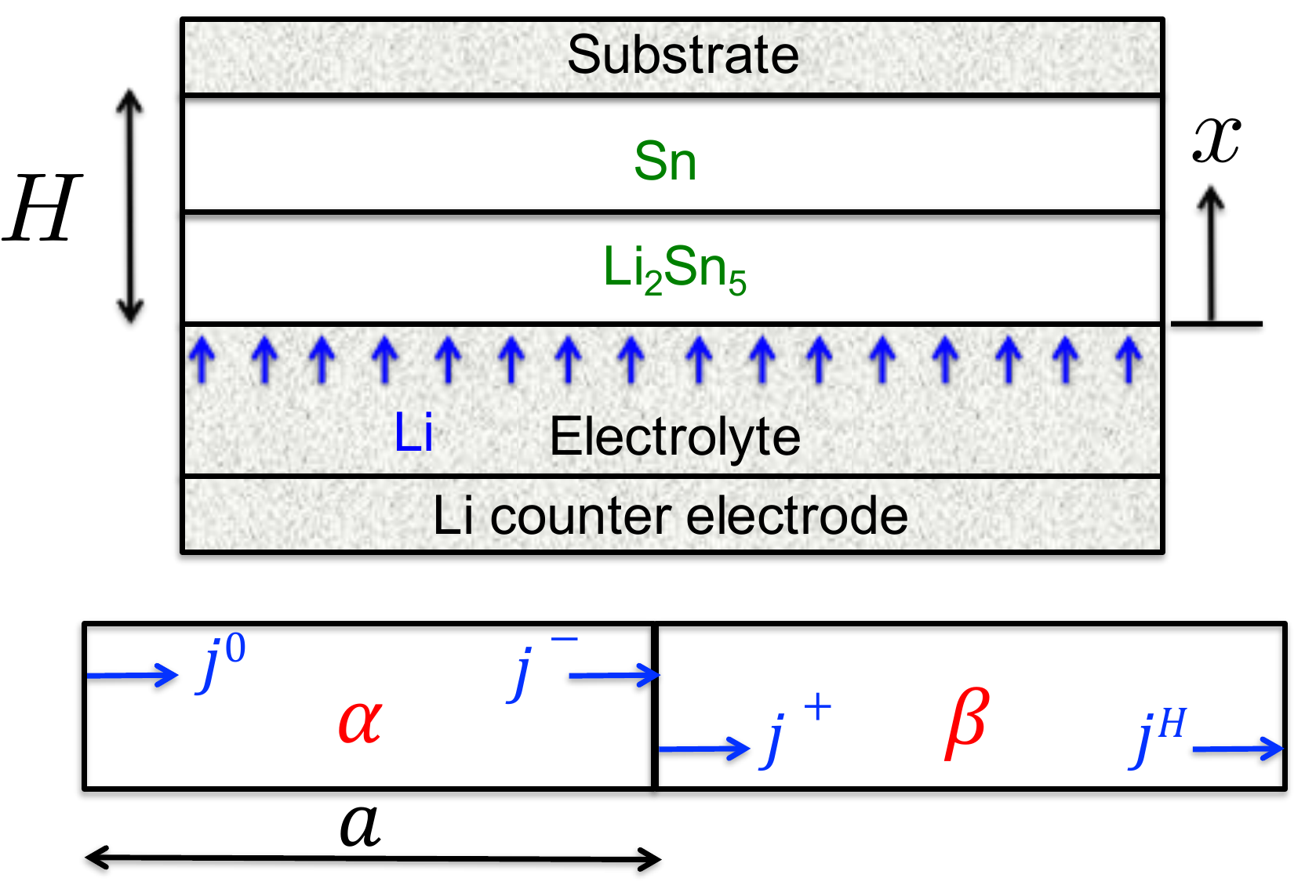}
\subcaption{}
\label{fig:electrodeSystemSchematic2}
\end{subfigure} 
\begin{subfigure}[c]{0.49\textwidth}
\includegraphics[width=\textwidth]{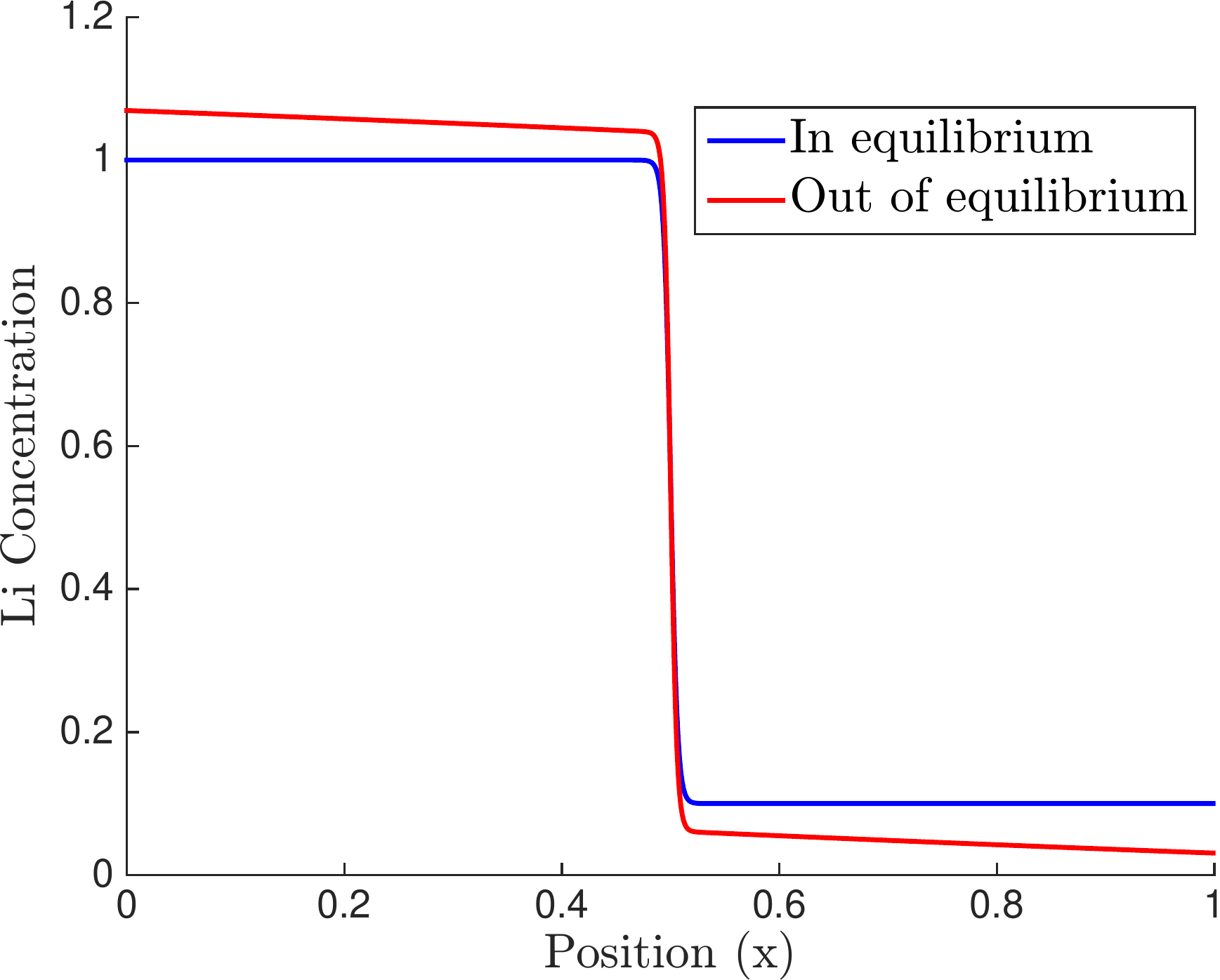}
\subcaption{}
\label{fig:concentrationProfileInOutEquilibrium}
\end{subfigure} 
\caption{(a) Schematic representing the one-dimensional modeling of the
thin-film experiments. (b) Li concentration profiles in the film at equilibrium
(blue) and out of equilibrium due to a flux of Li (red). In
Cahn-Hilliard models, concentration is also the phase variable.}
\label{fig:electrodeSystemSchematicTheoryConcentrationProfile}
\end{figure}

We are interested in modeling the behavior of half-cells with a Li-Sn thin film
electrode as illustrated in Figure \ref{fig:electrodeSystemSchematic2}. We
assume infinitely fast transport of Li ions in the electrolyte.  The  electrode
initially consists of the pure Sn phase. Upon the application of an external
electrical connection between the electrodes, Li is inserted into the Sn
electrode through an electrochemical reaction at the electrode/electrolyte
interface. We assume that Sn atoms are immobile and Li atoms diffuse through
the film and occupy interstitial positions. When the Li concentration at the
electrode surface reaches a critical value, the second phase nucleates and a
phase boundary begins propagating through the film. Stresses accompanying
lithiation and phase transformations in Li-Sn are of the order of a few MPa and
we assume they do not significantly affect Li diffusion or interface
propagation \cite{RN178}.

The Li transport, phase nucleation, and interface motion are modeled using a
one-dimensional (along the thickness of the film) version of the Cahn-Hilliard
equation.
\begin{equation}\label{eq:cahnHilliard1d} \frac{\partial c}{\partial t} =
\frac{\partial}{\partial x} M \frac{\partial \mu}{\partial x}, \quad \mu =
\frac{dG_0}{dc}-\kappa \frac{\partial^2 c}{\partial x^2}\end{equation}
where $c = \rho_{\text{Li}}/\rho_{\text{Sn}}$ (ratio of molar densities of
Li and Sn), $\mu$ is the chemical potential of Li, $M$ is a function
that characterizes the mobility of Li, $\kappa$ is a parameter related to the
interface-energy (determines the phase-boundary width), and $G_0(c)$ is the
homogeneous free-energy of Li-Sn. 

We consider a homogeneous free-energy of the form 
\begin{equation}\label{eq:doubleWellFreeEnergy}
G_0(c)=\mu^\text{eq}c+\frac{W}{2(\Delta c_0)^2}(c-c_0^\alpha)^2(c-c_0^\beta)^2
\end{equation}
where $c_0^\alpha$ and $c_0^\beta$ are the equilibrium concentrations of the
two phases, $\Delta c_0 = c_0^\beta - c_0^\alpha$, $\mu^\text{eq}$ is the
equilibrium chemical potential, and $W$ is free-energy curvature at the
equilibrium concentrations. The quartic double-well energy has been used in
many phase-field studies. The linear term $\mu^\text{eq}c$ is necessary to get
the right plateau voltage between the two phases (see Section
\ref{sec:calibration}). An advantage of this free-energy is that it is amenable
to perturbation analysis (Section \ref{subsec:perturbationAnalysis}). Figure
\ref{fig:freeEnergyAndDerivativeDoubleWell} shows the free-energy $G_0(c)$ and
$dG_0/dc$ for $c_0^\alpha = 0.1, c_0^\beta = 1, \mu^\text{eq} = 1, W = 50$. The
free-energy determines the equilibrium chemical potential, spinodal and
metastable regions, and the nucleation potentials for the phase transformations
(see \cite{RN177} for more).

\begin{figure}[h]
\centering
\begin{subfigure}[t]{0.49\textwidth}
\includegraphics[width=\textwidth]{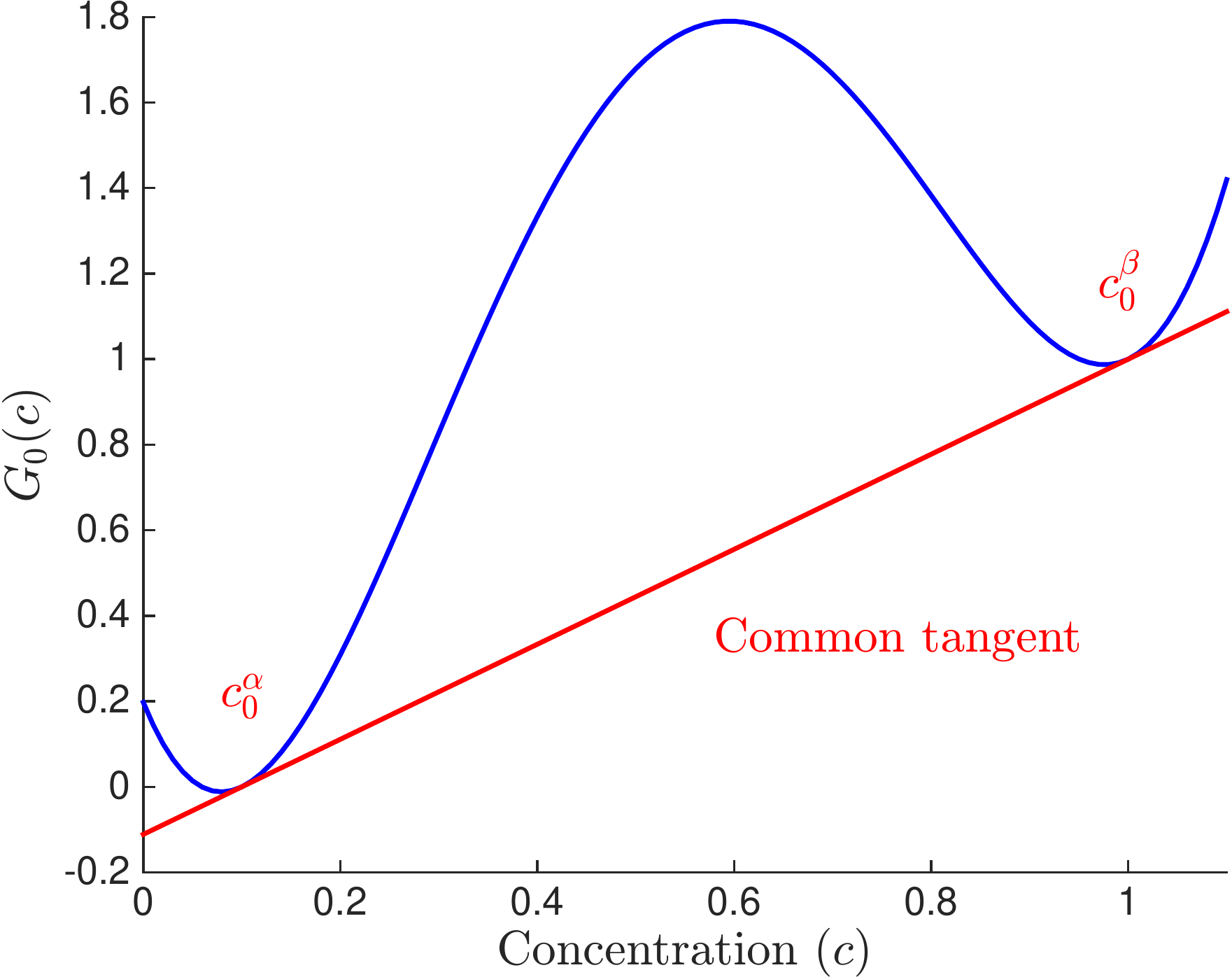}
\subcaption{}
\label{fig:doubleWellFreeEnergy}
\end{subfigure} 
\begin{subfigure}[t]{0.49\textwidth}
\includegraphics[width=\textwidth]{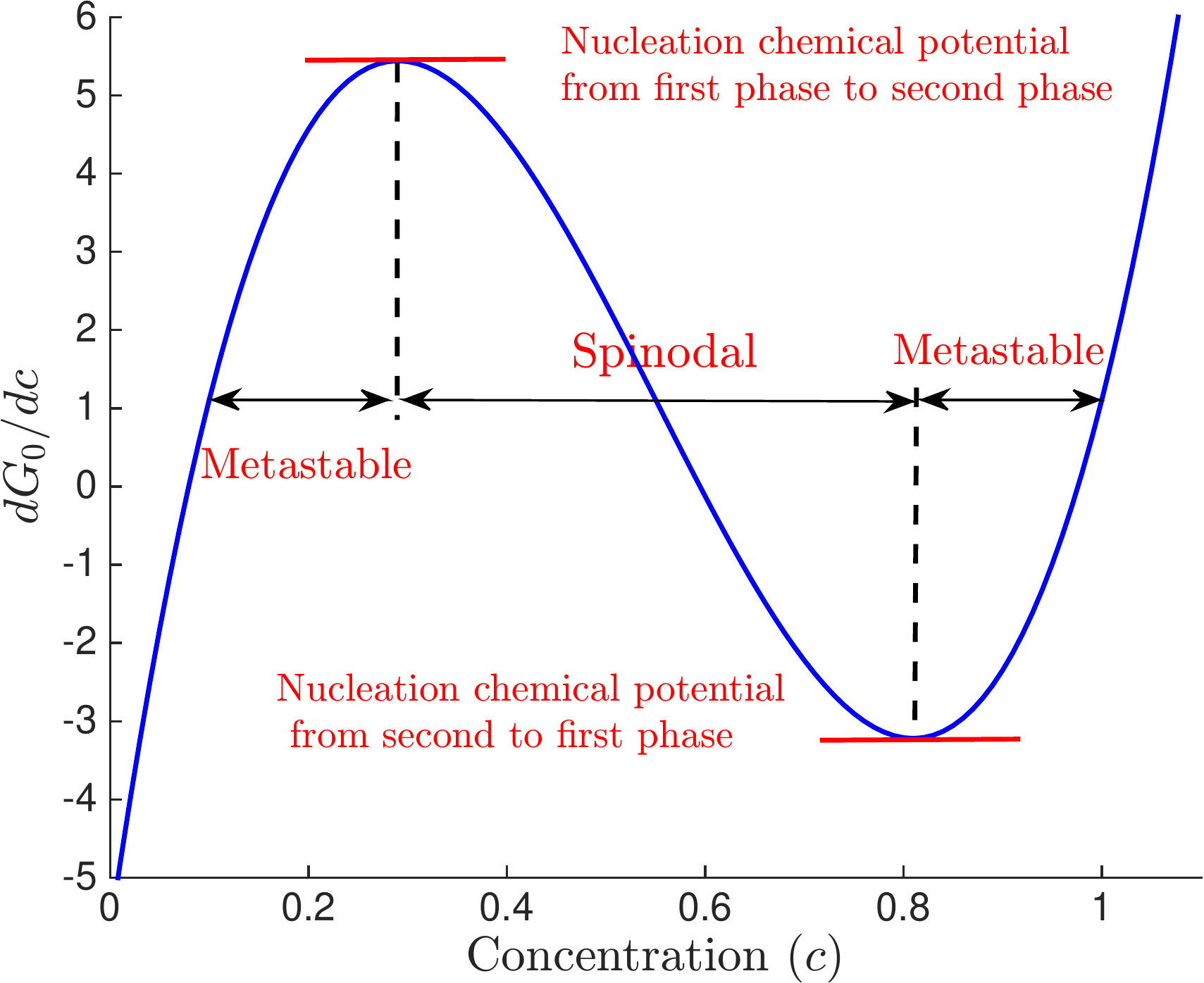}
\subcaption{}
\label{fig:freeEnergyDerivativeDoubleWell}
\end{subfigure} 
\caption{(a) Homogeneous free-energy $G_0$ (Equation
\ref{eq:doubleWellFreeEnergy}) and (b) its derivative $dG_0/dc$. Each energy
minimum corresponds to a stoichiometric phase. The common tangent for
equilibrium, the spinodal and metastable regions, and the nucleation potentials
for phase transformations are all shown.}
\label{fig:freeEnergyAndDerivativeDoubleWell}
\end{figure}

We assume the mobility $M$ in Equation \ref{eq:cahnHilliard1d} to depend on
the concentration gradient (see Sections \ref{sec:sharpInterface} and 
\ref{sec:modelBehavior} for more details)
\begin{equation}\label{eq:concentrationGradientKineticParameter}
M=\frac{M_0}{1+ \frac{\chi}{\Delta c_0} |\frac{\partial c}{\partial x}|}.
\end{equation}
Here, $M_0$ is the Li mobility in the bulk phase and
$\chi$ is a parameter that determines the interface mobility.  In the bulk,
concentration gradients are small, $M \approx M_0$ and the chemical
diffusivity of Li is given by $D = M_0 (d^2G_0/dc^2)$ (this can be seen by
linearizing the Cahn-Hilliard equation). $\chi = 0$ results in infinite
interface mobility while for a nonzero $\chi$, the interface mobility is finite
(see Section \ref{sec:modelBehavior} for details).

To predict the electrochemical response, we need to model the Li insertion
reaction, and possibly side-reactions. As a simple preliminary approximation,
we ignore side reactions and suppose that the electrode current is related to
the over-potential by the Butler-Volmer equation \cite{RN160}
\begin{equation}\label{eq:butlerVolmer}
i = i_0 (e^{\frac{\alpha \eta F}{RT}} - e^{-\frac{(1-\alpha) \eta F}{RT}}),
\end{equation}
where $i$ is the current density, $i_0$ is the exchange current density, which
may be a function of the Li concentration at the surface of the electrode (as
well as the Li ion concentration in the electrolyte), $\alpha$ is a
phenomenological constant (between 0 and 1, we assume the
anodic and cathodic transfer coefficients to be $\alpha$ and $1-\alpha$
respectively), $F$ is the Faraday constant, $R$ is the universal gas
constant, and $T$ is temperature. $\eta = V-U_0$ is the overpotential (the
difference between the externally applied potential V and the rest potential
$U_0$). The rest potential is related to the chemical potential at the surface
of the electrode by 
\begin{equation}\label{eq:referencePotential}
U_0 = -\frac{1}{F}(\mu^\theta+\mu)
\end{equation}
where $\mu^\theta$ is a (constant)
reference potential (its value depends on the choice of counter-electrode used
to define the electric potential). The electrode current $I$ is related to flux
$J$ at the electrode/electrolyte interface by $I = i A = -A\rho_{\text{Sn}}FJ$,
where $A$ is the area of the film.

The remaining boundary conditions are: $ \frac{\partial \mu}{\partial x} = 0$
at $x=H$ since the flux at the substrate is zero.  At the electrode/electrolyte
interface $x=0$, $-M \frac{\partial \mu}{\partial x} = J$. We also assume
$\frac{\partial c}{\partial x} = 0$ at $x = 0,H$.  \iftrue

\section{Sharp-interface model}\label{sec:sharpInterface}

We find it useful to study the sharp-interface limit of the governing equations
\ref{eq:cahnHilliard1d}-\ref{eq:concentrationGradientKineticParameter}. This
reveals the interface behavior and is helpful in finding model parameters from
experiments. We start by deriving a general energetic formulation of a
sharp-interface model that identifies the kinetic variables and their conjugate
forces for diffusion in the bulk and at the interface. Then we present the
results of a perturbration analysis of the governing quations
\ref{eq:cahnHilliard1d}-\ref{eq:concentrationGradientKineticParameter}.
Comparing the two identifies the kinetic relations for the interface implied by
the Cahn-Hilliard equations.

\subsection{A general sharp-interface model} \label{subsec:generalSharp}

Consider a thin film of thickness $H$ in which two phases $\alpha$ and $\beta$
are separated by an interface at $x=a$ (Figure
\ref{fig:electrodeSystemSchematic2}). The total energy
of the system is given by
\begin{equation}\label{eq:totalEnergySharpInterfaceModel} 
U = A \left[ \int_0^a G_\alpha(c(x)) \mathrm{d}x + \int_a^H
G_\beta(c(x)) \mathrm{d}x \right] 
\end{equation}
where $G_\alpha$  and $G_\beta$ are the free-energy densities of the two
phases, $A$ is the cross-sectional area, $a$ is the position of the interface,
and $H$ is the thickness of the film. We have ignored the surface energy since
it does not change with time. Differentiating Equation
\ref{eq:totalEnergySharpInterfaceModel} with respect to time, 
\begin{equation} 
\dot{U} =  A \left[ \int_0^a
\frac{dG_\alpha}{dc}\frac{\partial c}{\partial t} \mathrm{d}x + \int_a^H
\frac{dG_\beta}{dc}\frac{\partial c}{\partial t} \mathrm{d}x  +
\left\{G_\alpha(c^-)-G_\beta(c^+) \right\}\frac{da}{dt}\right]
\end{equation}
Using $\mu = dG/dc$, $\partial c/\partial t = -\partial j/\partial x$ (where
$j$ is the flux) and integrating by parts, we can write this as
\begin{equation} 
\dot{U} =  A \left[ \mu^0j^0 - \mu^H j^H + 
\int_0^H j \frac{\partial \mu}{\partial x} \mathrm{d}x  +
\mu^+j^+ - \mu^- j^- + 
\left\{G_\alpha(c^-)-G_\beta(c^+) \right\}\frac{da}{dt}\right]
\end{equation}
where $\mu^0, \mu^H, j^0,j^H$ are the chemical potentials and fluxes at $x=0$
and $x=H$ respectively. The first two terms capture energy flux at the
boundaries, the integral is the dissipation in the bulk due to diffusion, and
the rest correspond to dissipation at the interface. The superscripts +/– label
states just ahead and behind the interface. The interfacial dissipation is
\begin{equation} 
\dot{U}_{\text{int}} =  A \left[ \mu^+j^+ - \mu^- j^- +
\left\{G_\alpha(c^-)-G_\beta(c^+) \right\}\frac{da}{dt}\right]
\end{equation}
Using 
\begin{equation} \mu^+j^+ - \mu^- j^- = \frac{\mu^++\mu^-}{2}(j^+-j^-) + (\mu^+-\mu^-)
\frac{j^++j^-}{2} \end{equation}
and the conservation equation
\begin{equation}\frac{da}{dt} (c^+-c^-)  = j^+-j^-,  \end{equation}
the interfacial dissipation can be written as
\begin{equation} \dot{U}_{\text{int}} =  A \left[ \frac{\mu^+-\mu^-}{2}(j^++j^-) + \left\{
\frac{(\mu^++\mu^-)}{2} - \frac{G_\beta(c^+)-G_\alpha(c^-)}{c^+-c^-} \right\}
(j^+-j^-) \right] \end{equation}
This identifies the driving forces conjugate to the average flux and interface
velocity (the jump in flux is proportional to the interface
velocity). The interface kinetics is defined by prescribing the velocity and
average flux as a function of their driving forces. Assuming linear kinetics, 
\begin{equation}\label{eq:kineticRelations}
j^++j^- =-K_1\left(\mu^+-\mu^-\right), \quad j^+-j^- = -2K_2 \left\{
\frac{(\mu^++\mu^-)}{2} - \frac{G_\beta(c^+)-G_\alpha(c^-)}{c^+-c^-} \right\}.
\end{equation}
The conditions $K_1>0, K_2>0$ ensure that the interfacial dissipation is always
positive. At equilibrium, setting the driving forces to zero, we have
\begin{equation}
\mu^{eq+} = \mu^{eq-} = \frac{dG_\alpha}{dc}(c^{eq-})
=\frac{dG_\beta}{dc}(c^{eq+}) =
\frac{G_\beta(c^{eq+})-G_\alpha(c^{eq-})}{c^{eq+}-c^{eq-}}.
\end{equation}
At equilibrium, the chemical potential is continuous across the interface
and is given by the common tangent rule.
For small deviations about equilibrium, linearizing the kinetic relations
(Equation \ref{eq:kineticRelations}),
\begin{equation}\label{eq:kineticRelationsLinearized} 
j^++j^- = -K_1 \left(\delta\mu^+-\delta\mu^-\right), \quad 
j^+-j^- = -K_2 \left(\delta\mu^++\delta\mu^-\right)
\end{equation}
where $\delta\mu^+, \delta\mu^-$ are the deviations of the chemical potential
from their equilibrium value. The average flux at the interface is driven by
the jump in the chemical potential while the jump in the flux, which is 
proportional to the interface velocity, is driven by the sum of the deviations
of the chemical potential from equilibrium. The interfacial dissipation for
small deviations about equilibrium is
\begin{equation} \label{eq:interfacialDissipation}
\dot{U}_{\text{int}} =  -A \left[ K_1 \frac{(\delta\mu^+ - \delta\mu^-)^2}{2} +
K_2 \frac{(\delta\mu^+ + \delta\mu^-)^2}{2} \right] =
 -A \left[ \frac{(j^+ + j^-)^2}{2K_1 } + \frac{(j^+ - j^-)^2}{2K_2} \right]
\end{equation}
If flux in the bulk is proportional to the chemical potential gradient, \begin{equation}j =
-M \frac{\partial \mu}{\partial x}, \quad M > 0, \end{equation} bulk dissipation is
always positive. The bulk dissipation is 
\begin{equation}
\label{eq:bulkDissipation} \dot{U}_{\text{bulk}} =  A \int_0^H j \frac{\partial
\mu}{\partial x} \mathrm{d}x = -A \int_0^H \frac{j^2}{M} \mathrm{d}x.
\end{equation} 

\iftrue
\subsection{Perturbation analysis of Cahn-Hilliard equations}
\label{subsec:perturbationAnalysis}

With the above general formulation, let us turn our attention back to the Cahn-Hilliard
equations. Here, we borrow the results from the perturbation analysis of Langer
and Sekerka \cite{RN160} and refer the reader to that paper for details. Langer
and Sekerka studied the nonequilibrium interface behavior for small
perturbations about its equilibrium (of equations
\ref{eq:cahnHilliard1d}-\ref{eq:concentrationGradientKineticParameter}).  

When the interface is at equilibrium, the concentrations at its boundaries
reach the equilibrium values $c_0^\alpha,c_0^\beta$ given by the common tangent
rule (Figure \ref{fig:freeEnergyAndDerivativeDoubleWell}). When it is pushed
out of equilibrium, it is moving and/or there is a flux through it, the
concentrations deviate from $c_0^\alpha,c_0^\beta$ (Figure
\ref{fig:concentrationProfileInOutEquilibrium}). Langer and Sekerka showed that
the deviations as a function of the interface velocity and flux are 
\begin{equation} \delta c^- = -\frac{[c_0] \chi}{12 M_0 W}v + \frac{\chi}{4 M_0
W} (j^-+j^+), \quad \delta c^+ = -\frac{[c_0] \chi}{12 M_0 W}v  - \frac{\chi}{4
M_0 W} (j^-+j^+) \end{equation}
where $\delta c^-,\delta c^+$ are the concentration deviations from
equilibrium, $j^-,j^+$  are the fluxes at the interface boundaries, $v$ is the
interface velocity, and $[c_0]$ is the jump in the equilibrium concentration.
Using the jump conservation equation $v[c]=[j]$ and approximating $[c]$ by
$[c_0]$ for small deviations from equilibrium, the above equations can be
written as 
\begin{equation}\label{eq:perturbationResults} \delta c^- =
\frac{\chi}{3 M_0 W} j^- + \frac{\chi}{6 M_0 W} j^+, \quad \delta c^+ =
-\frac{\chi}{6 M_0 W} j^- - \frac{\chi}{3 M_0 W} j^+ \end{equation} For small
deviations about the equilibrium, $\delta \mu = W \delta c$.  Using this,
\begin{equation}\label{eq:perturbationResultsChemicalPotential} 
\delta \mu^- = \frac{\chi}{3 M_0} j^- + \frac{\chi}{6 M_0} j^+, \quad
\delta \mu^+ = -\frac{\chi}{6 M_0} j^- - \frac{\chi}{3 M_0} j^+ 
\end{equation}

In the sharp-interface limit, one can think of these equations as boundary
conditions at the interface. To determine the evolution of the system, the
diffusion equations in the bulk have to be solved coupled with the above
equations at the interface. 

These results allow us to interpret roles of $\chi$ and $M_0$ in Equation
\ref{eq:concentrationGradientKineticParameter}.  If $\chi = 0$, $M = M_0$
and Equation \ref{eq:cahnHilliard1d} reduces to the standard Cahn-Hilliard
equation; $\delta c^{+/-} = 0,\delta \mu^{+/-} = 0$ and in the sharp-interface
limit, the interface is always in local equilibrium. For a nonzero $\chi$,
$\delta c^{+/-} \neq 0,\delta \mu^{+/-} \neq 0$ and depend on the velocity and
the flux. The number $\chi$ thus characterizes the interface-mobility. 

\subsection{Kinetic relations in the Cahn-Hilliard model}

From the general model and the perturbation analysis, we can infer the kinetic
relations implied by the sharp-interface limit of the Cahn-Hilliard equations.  We can
write the results of the perturbation analysis (Equations 
\ref{eq:perturbationResultsChemicalPotential}) as:
\begin{equation}\label{eq:perturbationResultsChemicalPotentialInverted}
j^++j^- = -\frac{2M_0 }{\chi} (\delta \mu^+ - \delta \mu^-), \quad
j^+-j^- = -\frac{6 M_0 }{\chi} (\delta \mu^+ + \delta \mu^-).
\end{equation}
Comparing these with the kinetic relations (Equations
\ref{eq:kineticRelationsLinearized}), 
\begin{equation}\label{eq:cahnHilliardKineticParameters}
K_1 = \frac{2M_0}{\chi}, \quad K_2 = \frac{6M_0}{\chi}.
\end{equation}
This shows that the Cahn-Hilliard equations are a particular case of a more general class
of kinetic laws where $K_2 = 3K_1$. Further, for constant kinetic parameter
$(\chi \rightarrow 0)$, $K_1,K_2 \rightarrow \infty$ and the interfacial
dissipation goes to zero for any finite flux (Equation
\ref{eq:interfacialDissipation}). Thus the standard Cahn-Hilliard formulation implies
zero energy dissipation due to interface propagation.

\subsection{Steady diffusion}\label{subsec:steadyState}

Let us consider the case when the lithiation rate is small enough that bulk
diffusion is at steady state. The chemical potentials in the bulk on either
side of the phase boundary are linear. At the interface, the chemical
potentials are given by the perturbation analysis (Equation
\ref{eq:perturbationResultsChemicalPotential}). If $\mu^0$ and
$\mu^H$ are the chemical potentials at $x=0$ and $x=H$,
then the fluxes in the bulk are
\begin{equation} j^- = -M_0 \frac{\mu^--\mu^0}{x_{\text{int}}},  \quad
j^+ = -M_0 \frac{\mu^H-\mu^+}{H-x_{\text{int}}}. \end{equation}
Flux at the substrate is zero, $j^+= 0$.
Using the perturbation analysis solutions (Equation
\ref{eq:perturbationResultsChemicalPotential}), $\mu^- = \mu^\text{eq}+\delta
\mu^-$, and the above equations,
\begin{equation}
\mu^0-\mu^\text{eq} = \frac{x_{\text{int}} +\chi/3}{M_0}j^-
\end{equation}
This allows us to determine material parameters $M_0$ and $\chi$ from
electrochemical experiments.  Linearizing the Butler-Volmer equation (see
equations \ref{eq:butlerVolmer} and \ref{eq:referencePotential}) for small
currents, 
\begin{equation} 
-\rho_{\text{Sn}} F J = i = i_0 \frac{\eta F}{RT}  = i_0
\frac{V F+\mu^\theta+\mu^0}{RT} 
\end{equation}
where $J$ the flux at the surface. Using the above two equations,
\begin{equation}\label{eq:currentEvolutionGalvanostaticLithiation}
V = -\frac{\mu^\theta +\mu^\text{eq}}{F} + \left[
\frac{x_{\text{int}}}{F M_0 \rho_{\text{Sn}}} + \frac{\chi}{3F M_0
\rho_{\text{Sn}}} + \frac{RT}{i_0} \right] i
\end{equation}
In a galvanostatic experiment, $i$ is a constant. The voltage $V$ varies
linearly with the interface position $x_{\text{int}}$ (for given $i$) and the
current $i$ (for given $x_{\text{int}}$).  These can be used to determine the
kinetic parameter $M_0$ and the interface-mobility parameter $\chi$ if we know
the Butler-Volmer constant $i_0$ (see Section \ref{sec:calibration}). 

\fi
\fi 

\section{Model behavior}\label{sec:modelBehavior}

Let us now look at a few representative examples that elucidate the general
behavior of the modified Cahn-Hilliard model.  Using the film thickness $H$, a
typical relaxation time $t^* = 10$ hours (see Figure
\ref{fig:experimentalResults}), and $RT$ to nondimensionalize length, time, and
energy respectively, the governing equations take the form
\begin{equation}\label{eq:governingEquationsNondim}
\bar{\mu} = \frac{d\bar{G}_0}{dc}-\bar{\kappa} \frac{\partial^2 c}{\partial
\bar{x}^2}, \quad \frac{\partial c}{\partial \bar{t}} =
\frac{\partial}{\partial \bar{x}} \bar{M} \frac{\partial \bar{\mu}}{\partial
\bar{x}}, \quad
\bar{M}=\frac{\bar{M}_0}{1+ \frac{\bar{\chi}}{\Delta c_0} |\frac{\partial c}{\partial \bar{x}}|}
\end{equation}
\begin{equation} \frac{\partial \bar{\mu}}{\partial \bar{x}} = 0 \text{ at } \bar{x} =
1, \quad -\bar{M}\frac{\partial \bar{\mu}}{\partial \bar{x}} = \bar{J}
\text{ at } \bar{x} = 0, \quad \frac{\partial c}{\partial \bar{x}} = 0
\text{ at } \bar{x} = 0,1 \end{equation}

\begin{equation}
\bar{J} = \bar{i} = \bar{i}_0 e^{\alpha \bar{\eta}}(1 - e^{-\bar{\eta}}), \quad
\bar{\eta} = \bar{V}-\bar{U}_0, \quad \bar{U}_0 = -(\bar{\mu}^\theta+\bar{\mu}) 
\end{equation}
where 
\begin{equation}
\left[\bar{G}_0,\bar{\mu},\bar{\mu}^\theta\right] =
[G_0,\mu,\mu^\theta]\frac{1}{RT}, \quad \bar{\kappa} = \frac{\kappa}{RT H^2},
\quad \bar{M} = M \frac{RT t^*}{ H^2}, \end{equation} \begin{equation} \bar{J}
= J\frac{t^*}{H}, \quad \bar{I} = I\frac{t^*}{F
\rho_{\text{Sn}} A H},\quad [\bar{i},\bar{i}_0] = [i,i_0]\frac{t^*}{F
\rho_{\text{Sn}} H}, \quad \quad \left[\bar{\eta},\bar{V},\bar{U}_0\right] =
\left[\eta,V,U_0\right]\frac{F}{RT}.
\end{equation}

Equation \ref{eq:currentEvolutionGalvanostaticLithiation} after
nondimensionalization is
\begin{equation}\label{eq:voltageEvolutionGalvanostaticLithiationNondim}
\bar{V}  = - \bar{\mu}^\theta -\bar{\mu}^\text{eq} + \left[
\frac{\bar{x}_{\text{int}}}{\bar{M}_0} + \frac{\bar{\chi}}{3\bar{M}_0} +
\frac{1}{\bar{i}_0} \right]\bar{I}
\end{equation}

\subsection{Galvanostatic lithiation}

Let us first look at galvanostatic lithiation (see Figure
\ref{fig:electrodeSystemSchematic2} for a schematic of the system). Initially,
the Li concentration in the film is uniform ($c = 0$). We start inserting Li at
$\bar{x} = 0$ at a constant flux $\bar{J}$. The concentration gradually
increases and when it reaches the (lower) spinodal concentration at the
boundary $\bar{x} = 0$, the new phase nucleates. Upon further lithiation, the
phase boundary propagates into the film. The concentration profiles just before
nucleation and during phase propagation are shown in Figure
\ref{fig:phasePropagationGalvanostaticLithiationDelithiation}. During
delithiation, the low concentration phase nucleates at the boundary
$\bar{x}=0$. If the applied current is small enough that the concentration at
$\bar{x}=0$ does not reach the (higher) spinodal concentration, the existing
phase boundary is pulled toward $\bar{x}=0$.

\begin{figure}[h]
\centering
\begin{subfigure}[t]{0.49\textwidth}
\includegraphics[width=\textwidth]{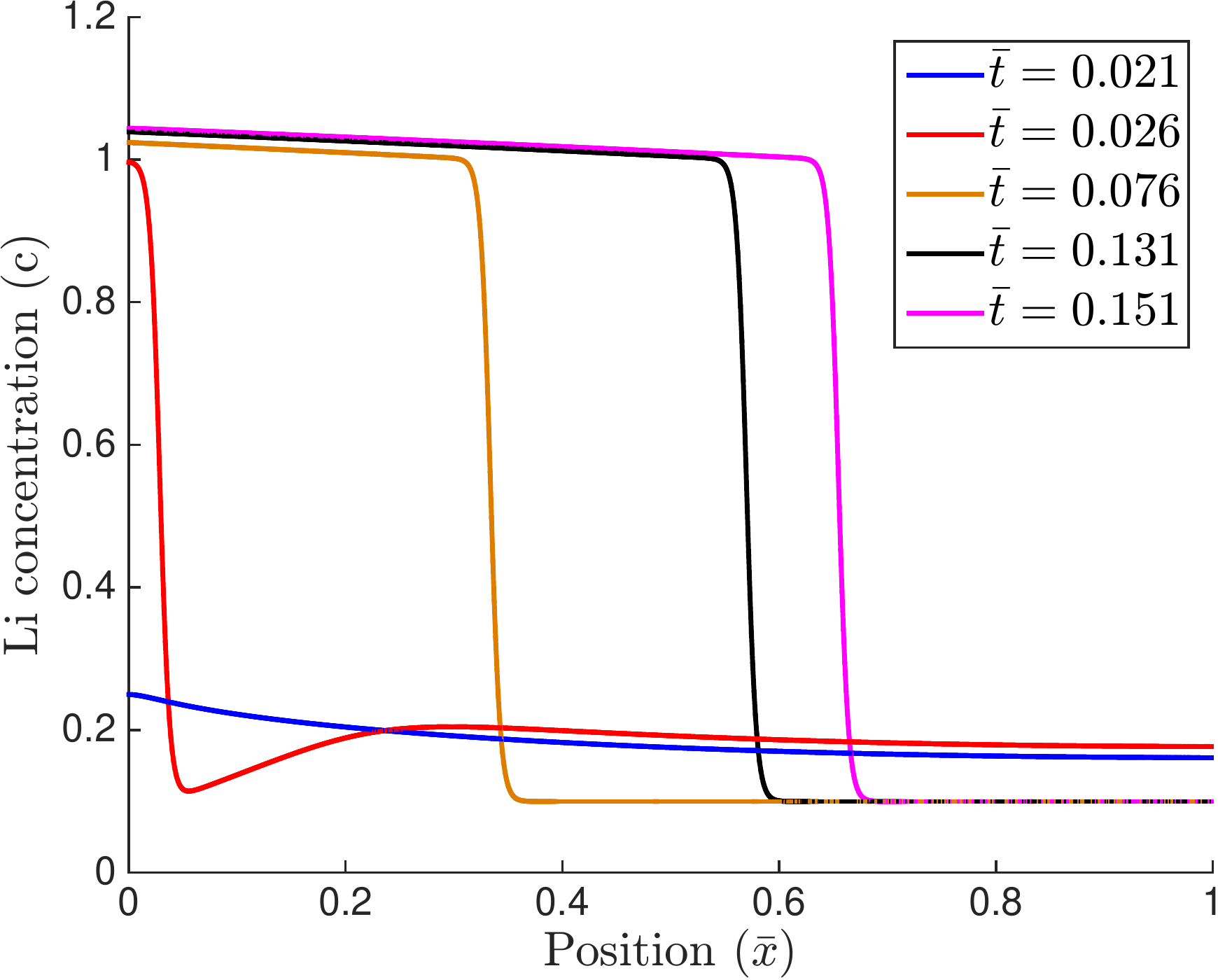}
\subcaption{}
\label{fig:phasePropagationGalvanostaticLithiation}
\end{subfigure} 
\begin{subfigure}[t]{0.49\textwidth}
\includegraphics[width=\textwidth]{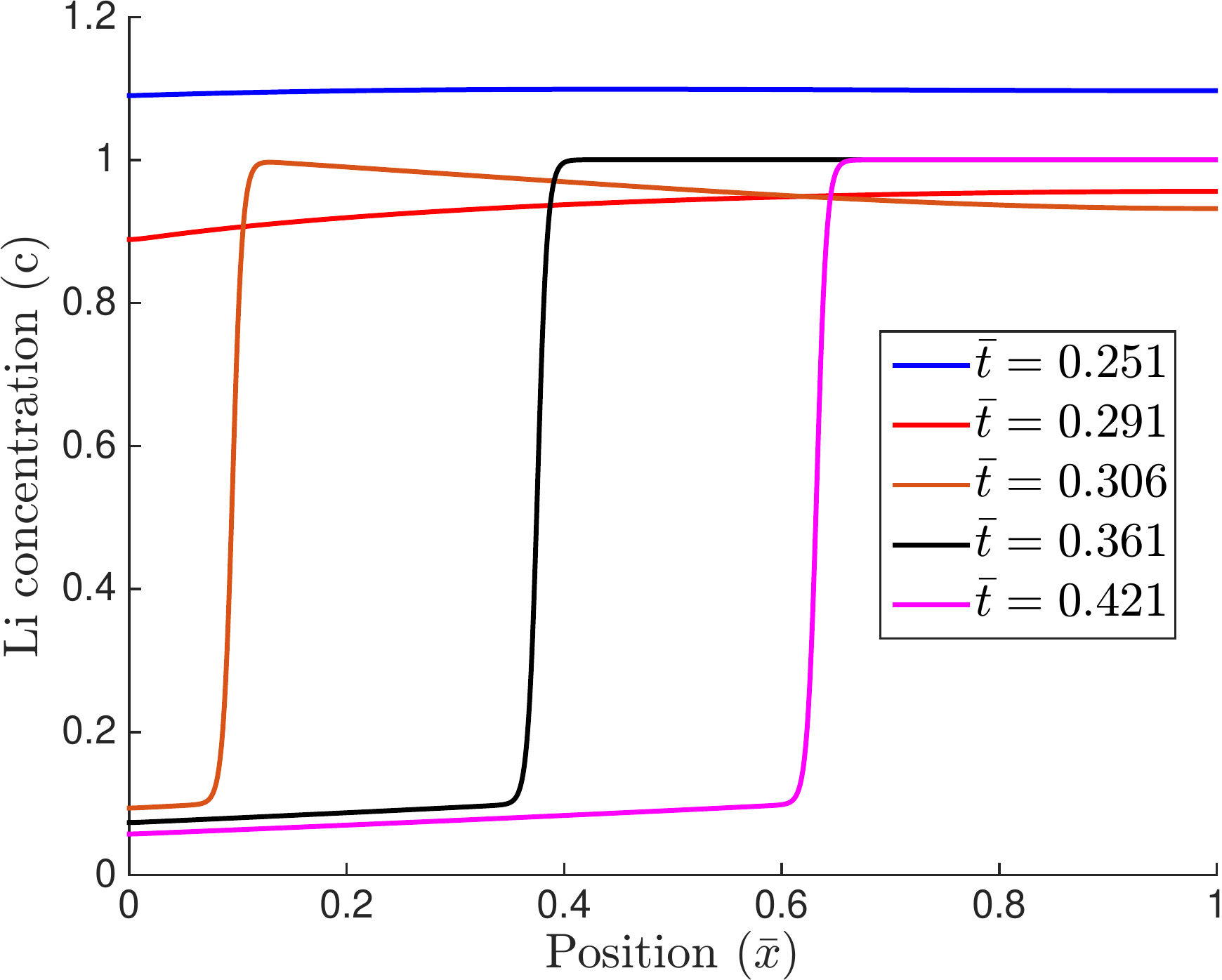}
\subcaption{}
\label{fig:phasePropagationGalvanostaticDelithiation}
\end{subfigure} 
\caption{Snapshots of concentration profile during galvanostatic (a)
lithiation and (b) delithiation. In both cases, a new phase nucleates at
$\bar{x}=0$ and propagates into the film. The nucleation occurs when the
concentration at $\bar{x}=0$ reaches that at the boundary of the spinodal
region (see Figure \ref{fig:freeEnergyAndDerivativeDoubleWell}).}
\label{fig:phasePropagationGalvanostaticLithiationDelithiation}
\end{figure}

The voltage evolution during galvanostatic lithiation is shown in Figure
\ref{fig:capacityVoltageCurrentDependence} for three different values of the
applied flux $\bar{J}$. Initially, as we start inserting Li, the voltage
decreases until nucleation of the new phase. During nucleation, the voltage
increases as the concentration at the boundary increases to that of the new
phase resulting in a characteristic bump. The slope of the voltage-capacity
during subsequent phase propagation increases with increasing flux. This is
because the slope of the chemical potential in the bulk is (at steady-state)
proportional to the flux $\bar{J}$, larger currents lead to larger chemical
potential change requiring lower voltages to keep inserting Li at the same
rate. The deviation of voltage from the equilibrium voltage (here
$\bar{V}^\text{eq} = -1$) during phase propagation also depends on the
Butler-Volmer parameter $\bar{i}_0$, here we have used $\bar{i}_0 = 1$.

The above described features of a voltage bump during nucleation and increasing
slope of voltage-capacity with increasing flux is observed in the experiments
as well. For example, Figure \ref{fig:capacityVoltageHuggins1999} shows
experimentally measured voltage evolution during galvanostatic
lithiation/delithiation for Li-Si/Li-Sn mixed-matrix electrode \cite{RN165}. 

\begin{figure}[h]
\centering
\begin{subfigure}[t]{0.52\textwidth}
\includegraphics[width=\textwidth]{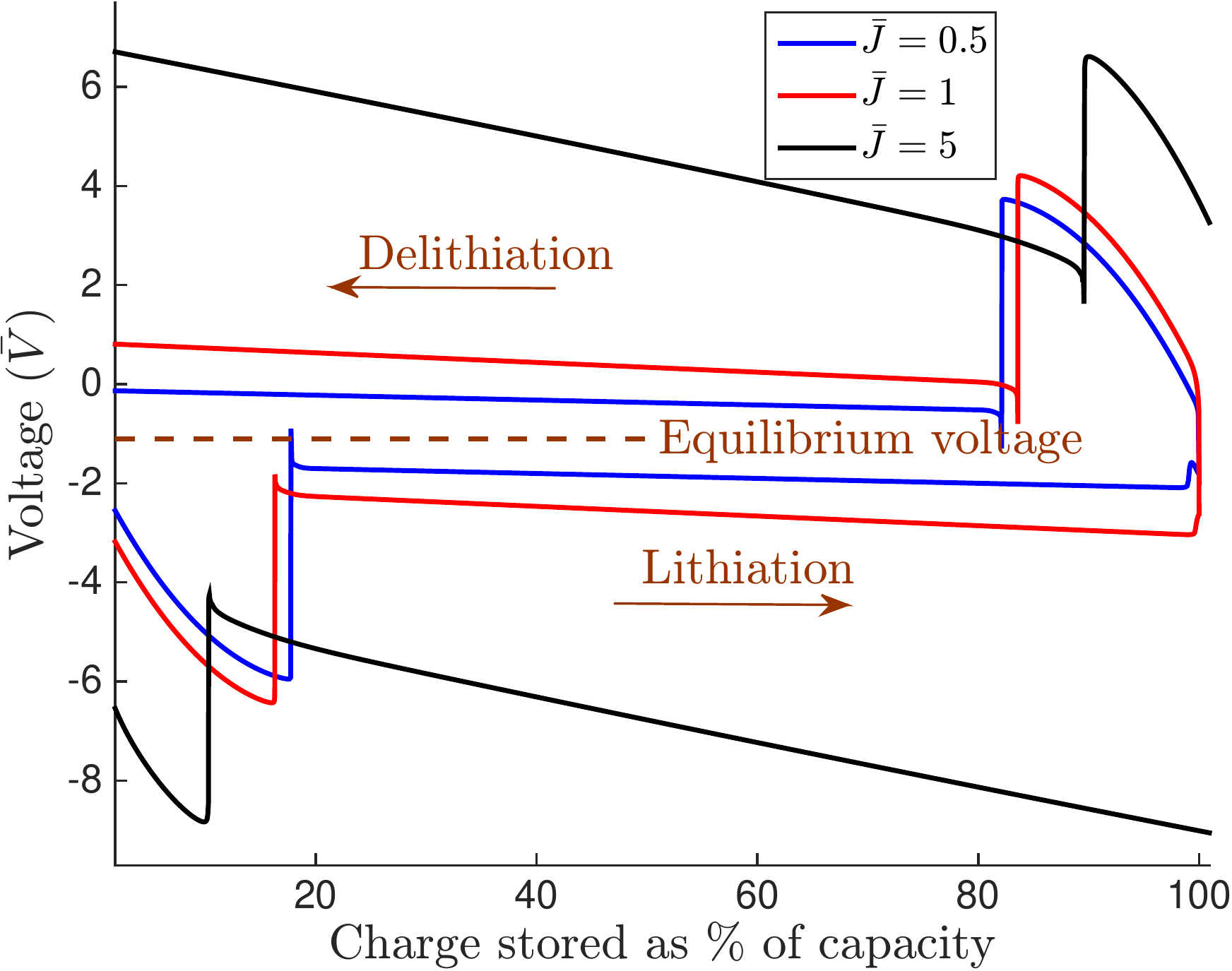}
\subcaption{}
\label{fig:capacityVoltageCurrentDependence}
\end{subfigure} 
\begin{subfigure}[t]{0.44\textwidth}
\includegraphics[width=\textwidth]{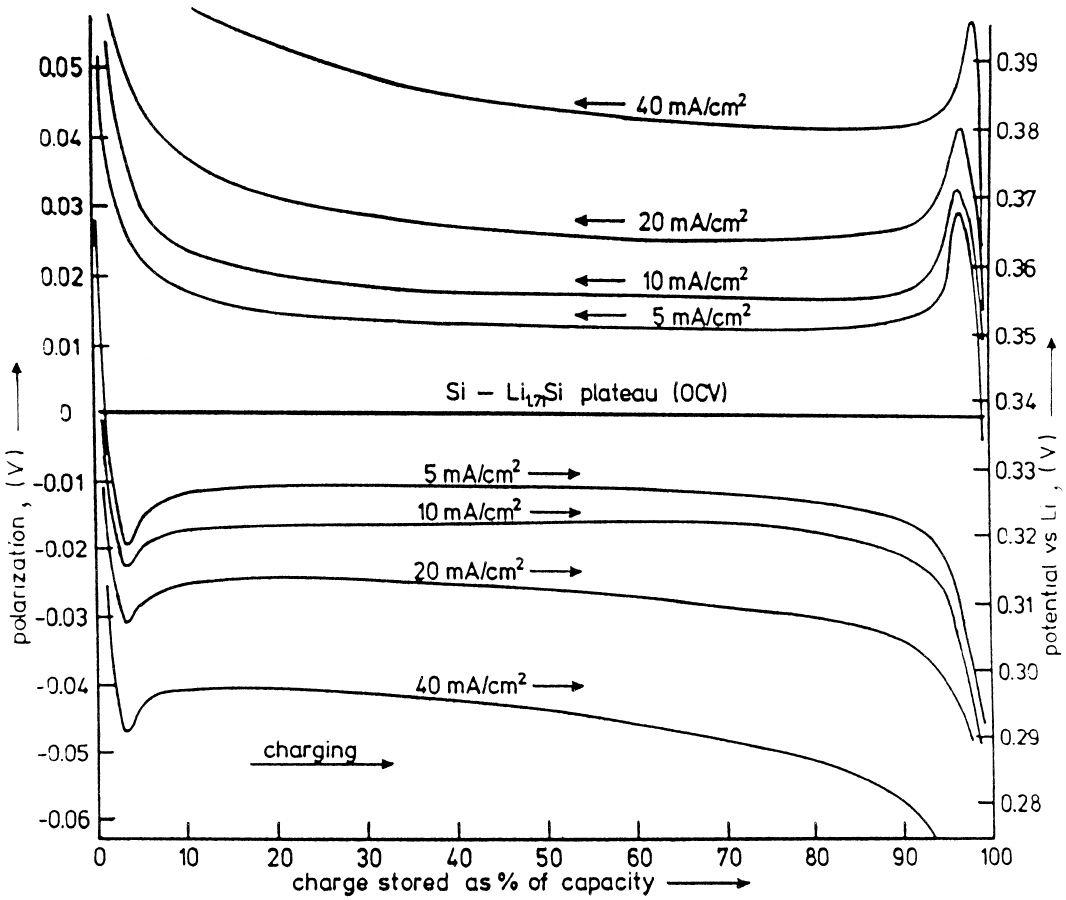}
\subcaption{}
\label{fig:capacityVoltageHuggins1999}
\end{subfigure} 
\caption{Voltage evolution during galvanostatic lithiation/delithiation in (a)
our simulations (b) experiments on a Li-Si/Li-Sn mixed-matrix electrode
\cite{RN165} (reproduced by permission of The Electrochemical Society). The
characteristic bump in the voltage corresponds to nucleation of a phase. The
slope during phase propagation increases with increasing current due to
diffusion-limited lithiation.}
\end{figure}

\subsection{Potentiostatic lithiation}

Potentiostatic lithiation is very useful in determining many material
properties such as the free-energy, diffusivity, and interface-mobility. A
representative potentiostatic lithiation simulation is shown in Figure
\ref{fig:chemicalPotentialEvolutionPotentiostaticLithiation}. We start with a
thin film in the first phase, the Li concentration is $c=0$. We then
instantaneously lower the voltage to $\bar{V} = \bar{V}_1$.  

After the voltage step, as we start inserting Li, the concentration and
chemical potential increase in the film and the current starts to drop. If the
concentration at the boundary reaches the (lower) spinodal concentration
(Figure \ref{fig:freeEnergyAndDerivativeDoubleWell}), a new phase nucleates and
propagates into the film. Three snapshots of chemical potential in the film
during phase propagation are shown in Figure
\ref{fig:chemicalPotentialEvolutionPotentiostaticLithiationZeroChi}. All the
chemical potential drop across the film happens only within the first phase. In
particular, the chemical potential is continuous across the phase-boundary and
its value there is equal to the equilibrium chemical potential (which is 1
here). This can also be seen from the perturbation analysis (Equation
\ref{eq:perturbationResultsChemicalPotential}) where $\bar{\chi} = 0 \implies
\delta \bar{\mu} = 0$. This demonstrates that in the standard Cahn-Hilliard model, the
interface has infinite mobility and is always in local equilibrium.

\begin{figure}[H]
\centering
\begin{subfigure}[t]{0.49\textwidth}
\includegraphics[width=\textwidth]{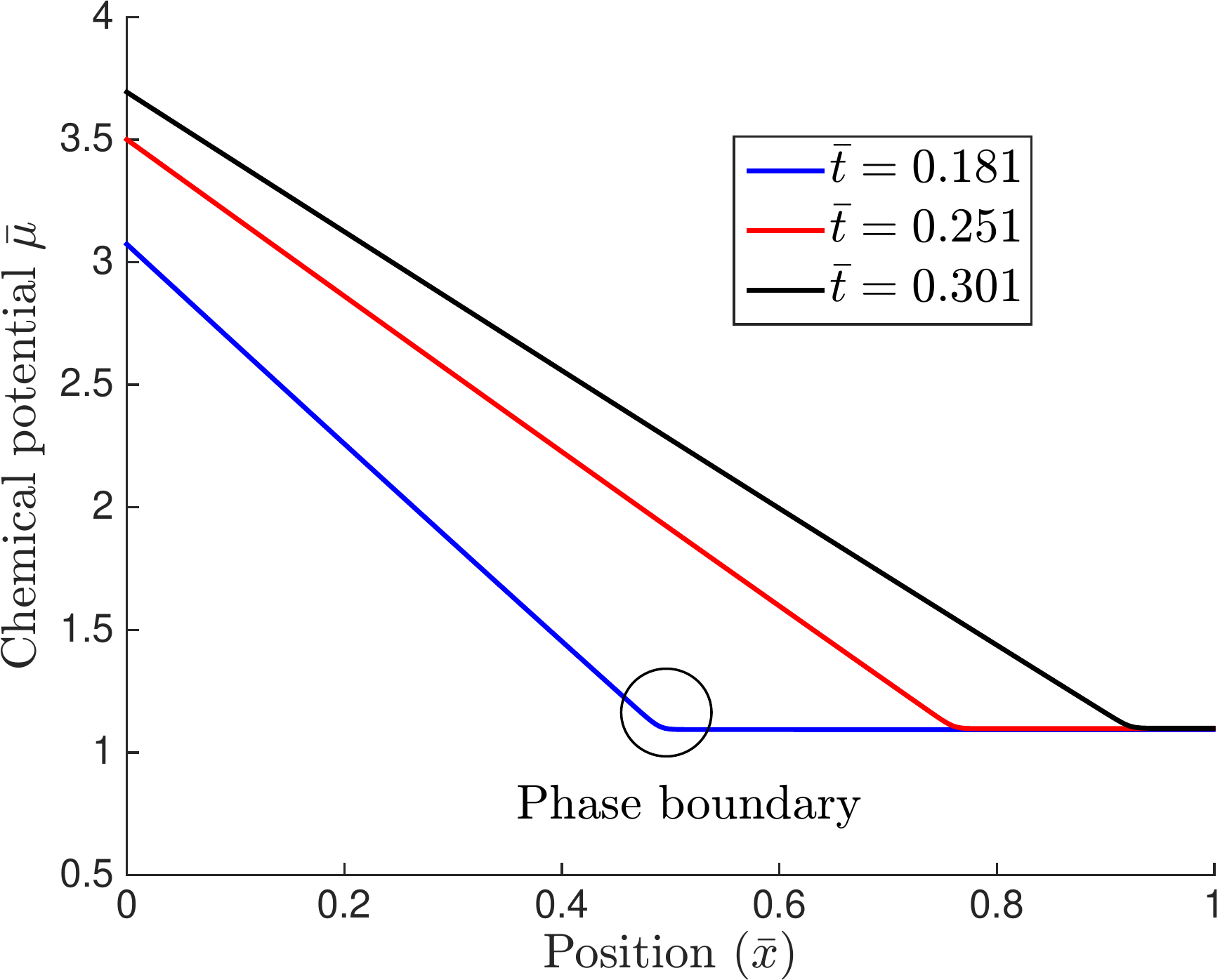}
\subcaption{}
\label{fig:chemicalPotentialEvolutionPotentiostaticLithiationZeroChi}
\end{subfigure} 
\begin{subfigure}[t]{0.49\textwidth}
\includegraphics[width=\textwidth]{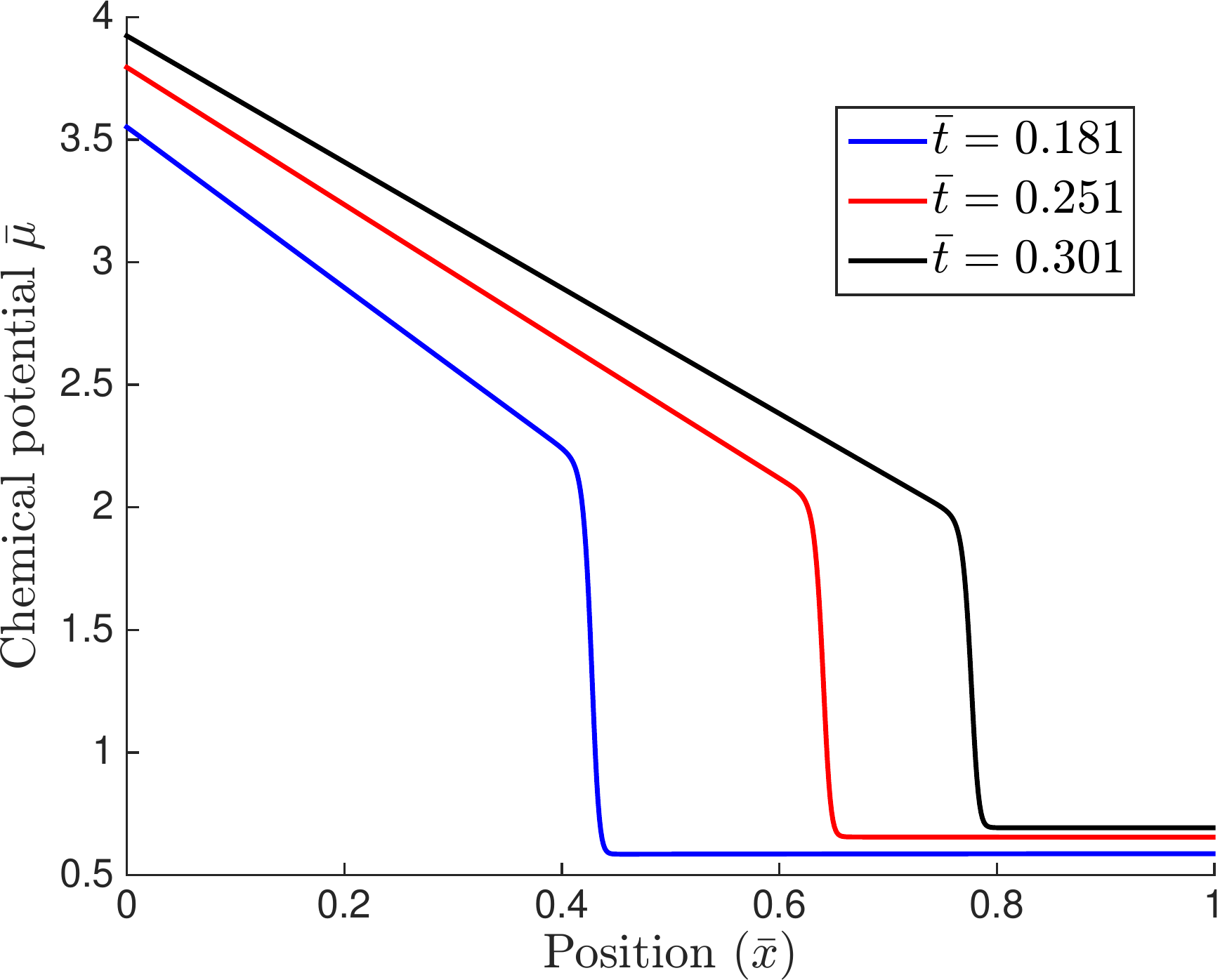}
\subcaption{}
\label{fig:chemicalPotentialEvolutionPotentiostaticLithiationNonzeroChi}
\end{subfigure} 
\caption{Three snapshots of chemical potential during phase propagation. (a)
$\bar{\chi} = 0$, all the chemical potential drop across the film happens
across the first phase. The jump in potential across the interface is zero and
the potential there is equal to the equilibrium potential. This demonstrates
that the interface in standard Cahn-Hilliard equations is always in local
equilibrium. (b) $\bar{\chi} = 0$, the chemical potential is now discontinuous
across the phase boundary. The interface is out of equilibrium, the deviation
of chemical potential from its equilibrium value acting as the driving force
conjugate to interface velocity. With increasing $\bar{\chi}$, the interface
becomes more sluggish, its velocity decreases, and it is pushed more out of
equilibrium.}
\label{fig:chemicalPotentialEvolutionPotentiostaticLithiation}
\end{figure}

A concentration-gradient dependent Li mobility (nonzero $\bar{\chi}$ in
Equation \ref{eq:concentrationGradientKineticParameter}) gives the interface a
finite mobility. We repeat the above potentiostatic lithiation simulation but
change $\bar{\chi}$ from 0 to 1. Figure
\ref{fig:chemicalPotentialEvolutionPotentiostaticLithiationNonzeroChi} shows
again, three snapshots of chemical potential. Comparing with Figure
\ref{fig:chemicalPotentialEvolutionPotentiostaticLithiationZeroChi}, we observe
two things. First, the chemical potential is now discontinuous across the
interface. Behind the interface, $\mu >\mu^\text{eq} $ while ahead of it $\mu <
\mu^\text{eq}$. As we saw earlier, the deviation of the chemical potential from
its equilibrium value acts as the driving force conjugate to the interface
velocity (Equation \ref{eq:perturbationResultsChemicalPotentialInverted}). The
magnitude of the potential jump across the phase-boundary decreases during
propagation since the current (flux) decreases (Equation
\ref{eq:perturbationResultsChemicalPotential}). Second, as $\bar{\chi}$
increases, lithiation gradually becomes interface limited and the phase
propagation velocity decreases. In the limiting case $\bar{\chi} \to \infty$,
the interface is immobile and all the diffusing Li accumulates behind the
phase-boundary. Thus, $\bar{\chi} = 0$ and $\bar{\chi} \to \infty$ are limits
of infinite mobility and immobility for the interface. This gives us the
ability to model both diffusion-limited and interface-limited processes as
opposed to the standard Cahn-Hilliard which can model only diffusion-limited lithiation.

There are three timescales: the surface-reaction timescale, the bulk-diffusion
timescale, and the interface timescale. Apart from other factors like the film
width $H$, the three are determined by $i_0, M_0, \chi$ respectively. We now
turn our attention to finding these numbers for the Li-Sn system.

\fi

\iftrue 
\section{Comparison of theory and experiment} \label{sec:calibration}

We have to determine the following parameters from experiments: the homogeneous
free-energy derivative $dG_0/dc$, the Butler-Volmer parameter $i_0$, the
kinetic parameter $M_0$, and  the interface-mobility parameter $\chi$. In
calibrating these parameters, it is important to determine how much of the
measured current in the experiment goes into the electrode and how much goes
into SEI formation.  To estimate this, we assume that all of the Li in the
first 20 hours (during which the potential is held at 0.8 V) goes into SEI and
use an exponential fit for the last few hours of current evolution in this
period. We assume that subsequent SEI growth follows this decay up to a
constant residual value (equal to the residual current we observe in the later
stages in our experiments),
\begin{equation}
I(t) = \begin{cases} I_0 e^{-(t-t_\text{initial})/t_\text{decay}}, & \mbox{if } t \leq t_\text{residual}, \\
                  I_{\text{residual}}, & \mbox{otherwise }, \end{cases}
\end{equation}
where the values of $I_0, t_\text{initial}$ depend on the initial time from
which the fit is made. $t_\text{residual}$ is the time at which $I_0
e^{-(t-t_\text{initial})/t_\text{decay}}$ reaches the residual current
$I_\text{residual}$. Typical value of $t_\text{decay}$ in our experiments, the
decay time for the SEI current, is about 30 hours. The SEI current at the end
of 20 hours is of the order of 0.1 $\mu$A cm$^{-2}$. The residual current
$I_{\text{residual}}$ observed is about 0.005 $\mu$A cm$^{-2}$.  This is
significantly smaller than the currents used in our calibration.  From the fit,
we calculate the total charge loss to SEI (over the timescale of our
experiments) to be about 0.05 C cm$^{-2}$ which is the same as that reported by
Bucci et al. \cite{RN53} (although that is for Si).

\subsection{Free-energy}
The homogeneous free-energy derivative $dG_0/dc$ is found from the steady-steady
voltage vs charge in a PITT experiment (Figure
\ref{fig:timeVoltageCurrentExperimentalPlotFirstLithiation}). After each
voltage step, we allow the current to drop to a prescribed value (0.05 mA g$^{-1}$,
about C/1800). From the total charge, we calculate the equilibrium Li
concentration at this voltage (these are preliminary results,
an alternative method would be to let the system go to OCV and use the voltage
corresponding to this state. In light of this, the numbers in Table
\ref{table:steadyStateVoltageConcentration} should be regarded only
qualitatively. We have not used these numbers in calibrating our free-energy).
Table \ref{table:steadyStateVoltageConcentration} shows the results of such a
calculation for the Sn phase. $c = 0$ and $c=0.4$ correspond to pure Sn and the
first phase Li$_2$Sn$_5$. The chemical potential can then be determined by
setting the overpotential to zero in the Butler-Volmer equation.
\begin{equation}
V = -\frac{1}{F}\left(\mu^\theta+\mu\right) = -U^\theta-\frac{1}{F} \frac{dG_0}{dc}
\implies \frac{dG_0}{dc} = -F\left( V + U^\theta \right),
\end{equation}
where $U^\theta = 2.75$ V is the open-circuit voltage with respect to
Li/Li$^+$.

The free-energy parameters (Equation \ref{eq:doubleWellFreeEnergy})  are
determined to satisfy the following conditions:
\begin{itemize}
\item Energy minima $(dG_0/dc = 0)$ at $c=0$ and $c= 0.4$ since these
correspond to stoichiometric phases of Sn and Li$_2$Sn$_5$ (Figure
\ref{fig:doubleWellFreeEnergy}). The curvatures at these concentrations
must be positive since they are energy minima $(d^2G_0/dc^2 > 0)$.
\item Plateau voltage between the first two phases is 0.75 V.
\item The (homogeneous) nucleation voltage for Sn$\rightarrow$Li$_2$Sn$_5$
transformation is around 0.7 V. With the first two constraints, the double-well
free-energy predicts a nucleation potential significantly larger than that in
experiment. Thus, we minimize the nucleation potential (this corresponds to
the limiting condition of the $d^2G_0/dc^2 = 0$ at $c = 0.4$).
\end{itemize}
The best-fit parameters based on the above conditions are:
\begin{equation}\label{eq:doubleWellFreeEnergyBestFitParameters}
c_0^\alpha = 0.0346, c_0^\beta = 0.4853, \mu^\text{eq} = 0.193
\text{ MJ mol}^{-1}, W = 4.57 \text{ MJ mol}^{-1}.
\end{equation}
The free-energy derivative based on the above numbers and that determined from
experiment are shown in Figure
\ref{fig:homogenousFreeEnergyExperimentsAfterSEICorrection}. The double-well
energy predicts a nucleation potential about twice as large as that observed in
experiment. This suggests that free-energies such as Equation
\ref{eq:doubleWellFreeEnergy}, though convenient for analysis, may not be
(quantitatively) good in capturing the concentration dependence of the
free-energy. Predictive phase-field models require a better way of
incorporating free-energy determined from experiments.

\begin{table}[h]
\centering
\begin{tabular}{|c|c|c|}
\hline
 Voltage(V)  &  Concentration \\ \hline
 2.7535 & 0 \\ \hline
 0.78 & 0.1501 \\ \hline
 0.76 & 0.1931 \\ \hline
 0.74 & 0.2033 \\ \hline
 0.72 & 0.2131 \\ \hline
 2.7535 & 0.4 \\ \hline
\end{tabular}
\caption{Voltage vs concentration at steady-state during PITT.
The concentration is calculated from the total charge after correcting for loss
to SEI.}
\label{table:steadyStateVoltageConcentration}
\end{table}

\begin{figure}[h]
\centering
\includegraphics[scale=0.5]{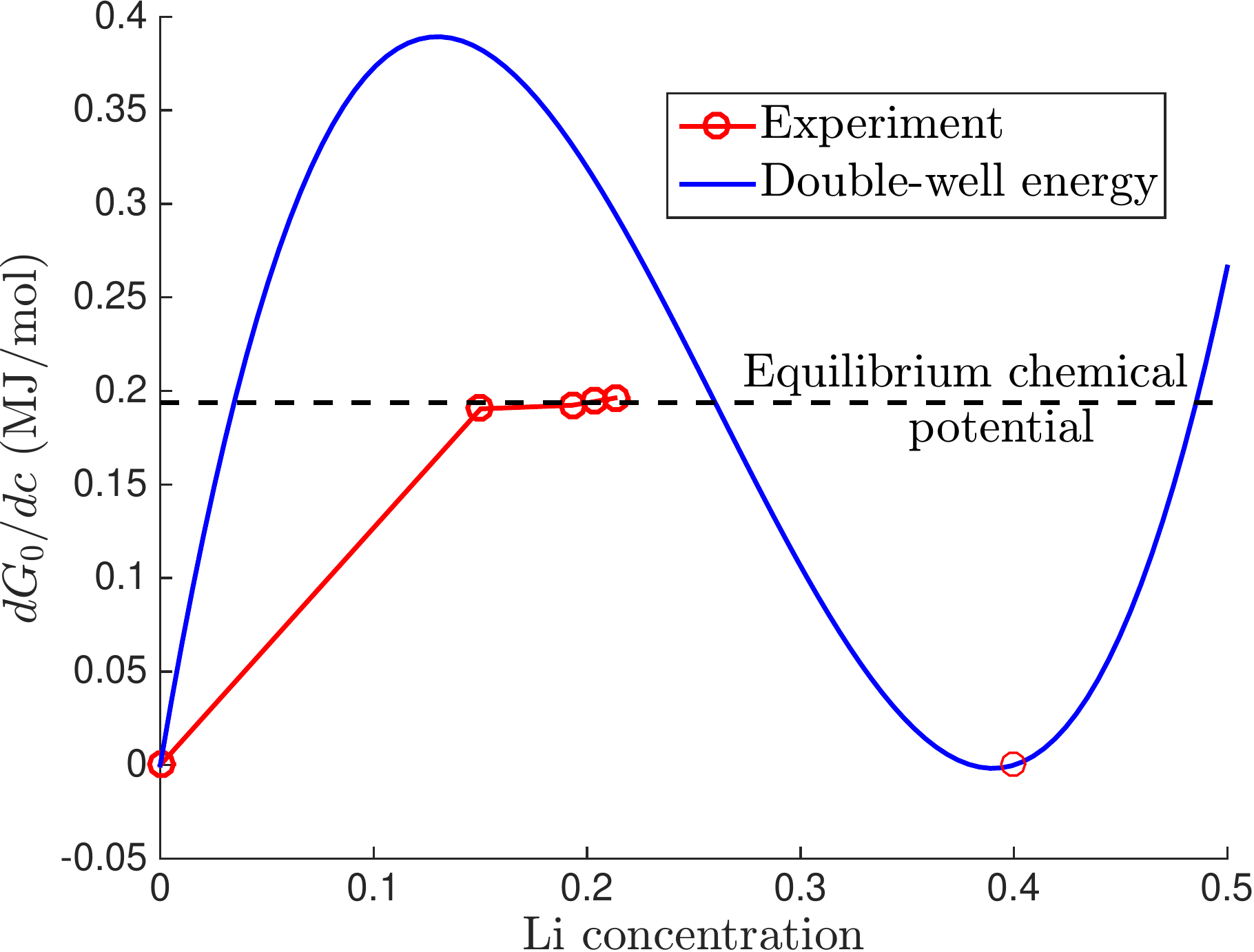}
\caption{Homogeneous free-energy derivative from steady-state voltage vs charge
measurements (Table \ref{table:steadyStateVoltageConcentration}) and using the
double-well free-energy (Equations \ref{eq:doubleWellFreeEnergy} and
\ref{eq:doubleWellFreeEnergyBestFitParameters}). The double-well predicts a
nucleation potential about twice that found in experiment. Outside of
nucleation however, the predictions match the experiments well since these are
determined by the free-energy near the equilibrium concentrations (close to
$c=0.0346$ and $c=0.4853$).}
\label{fig:homogenousFreeEnergyExperimentsAfterSEICorrection}
\end{figure}

\subsection{Diffusivity and exchange-current density of Sn}

To find the diffusivity $D$ and the exchange-current density $i_0$, we use the
current evolution after a voltage step (see inset in Figure
\ref{fig:timeVoltageCurrentExperimentalPlotFirstLithiation}) in PITT
experiments on films of Sn. We use two different methods, the first using the
Cottrell equation \cite{RN176} and the second based on the paper by Li et al.
\cite{RN168}. The Cottrell solution gives the short-time ($Dt/H^2 \ll 1$)
current response following a step change in voltage starting from a uniform
initial state assuming that the rate-limiting process is diffusion
\cite{RN168}:
\begin{equation}\label{eq:shortTimeCottrell}
I(t) = \frac{Q}{H} \sqrt{\frac{D}{\pi t}},
\end{equation}
where $I(t)$ is the current, $Q$ is the total charge transferred in the voltage
step, $H$ is the film thickness, and $D$ is the diffusivity.  Thus, this gives
us only the diffusivity. Li et al. \cite{RN168} derive analytical solutions for
current evolution considering a finite surface-reaction rate. The short-time
($Dt/H^2 \ll 1$) solution is given by
\begin{equation}\label{eq:shortTimeVerbrugge}
I(t) = \frac{DQB}{H} \exp{\left(B^2Dt\right)} \erf{\left(B \sqrt{Dt}\right)},
\end{equation}
where $B = -i_0(d^2G_0/dc^2)/(\rho_{\text{Sn}}FDRT)$. From this, we can
determine both the diffusivity and the Butler-Volmer parameter $i_0$. When the
surface-reaction is much faster than diffusion, $B \gg 1$, Equation
\ref{eq:shortTimeVerbrugge} reduces to Equation \ref{eq:shortTimeCottrell}.
Equations \ref{eq:shortTimeCottrell} and \ref{eq:shortTimeVerbrugge} are used
to fit the current evolution after a voltage step during PITT. Figure
\ref{fig:shortTimeCurrentVsTimeFit_FirstLithiation} shows one typical fit for
each method. The kinetic parameter $M_0$ is found from the diffusivity using $D
= M_0 (d^2G_0/dc^2)$ with the free-energy curvature calculated from the steady
state voltage vs charge measurements (Table
\ref{table:steadyStateVoltageConcentration}). Table
\ref{table:diffusivityCalibrationSn} shows values of the diffusivity, $M_0,
i_0$ from such fits at different voltages (for the Sn phase). 

\begin{figure}[h]
\centering
\begin{subfigure}[t]{0.49\textwidth}
\includegraphics[width=\textwidth]{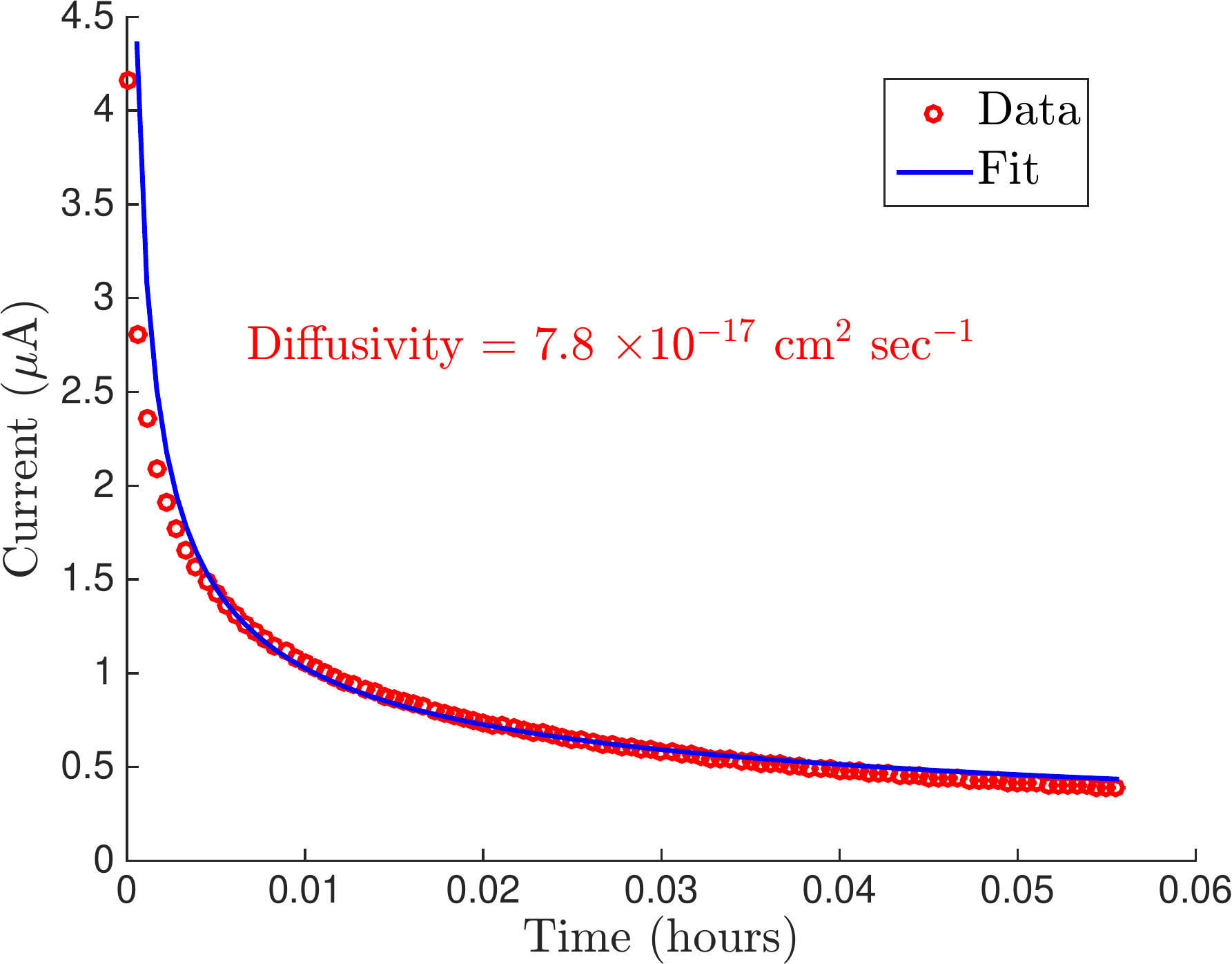}
\subcaption{}.
\label{fig:shortTimeCurrentVsTimeCottrellFit_FirstLithiation_074V}
\end{subfigure} 
\begin{subfigure}[t]{0.49\textwidth}
\includegraphics[width=\textwidth]{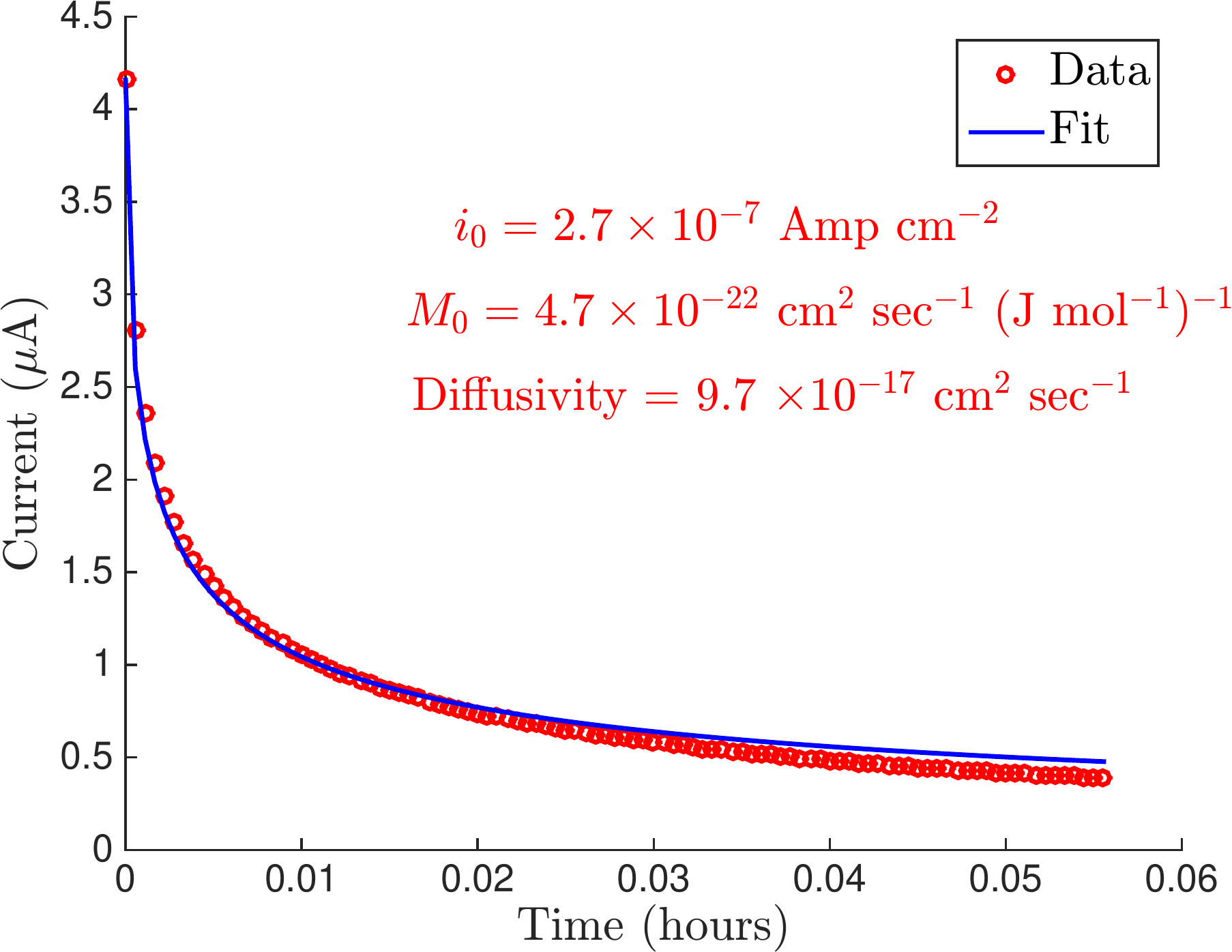}
\subcaption{}.
\label{fig:shortTimeCurrentVsTimeFit_FirstLithiation_074V}
\end{subfigure} 
\caption{Current vs time following a voltage step in PITT experiments. Fits
based on (a) Cottrell equation (b) Li et al.  \cite{RN168}. Results of such
fits done at different voltages are tabulated in Table
\ref{table:diffusivityCalibrationSn}.}
\label{fig:shortTimeCurrentVsTimeFit_FirstLithiation}
\end{figure}

\begin{table}[H]
\centering
\begin{tabular}{|c|c|c|c|c|c|}
\hline
  &  \multicolumn{3}{c|}{Li et al. based \cite{RN168}} &  \multicolumn{2}{c|}{Cottrell based} \\ \hline
	\specialcell{Voltage \\V} &  \specialcell{$i_0$ \\ $\mu$A cm$^{-2}$} & \specialcell{$M_0 $ \\$10^{-21}$cm$^2$sec$^{-1}$ (J mol$^{-1}$)$^{-1}$} & \specialcell{Diffusivity \\ $10^{-16}$ cm$^2$sec${^-1}$}
& \specialcell{Diffusivity \\ $10^{-16}$ cm$^2$sec${^-1}$} & \specialcell{Slope in  \\ log-log plot} \\ \hline
		0.72 & 0.29  & 0.76 &  1.50 & 1.56 &  -0.46 \\ \hline
    0.74 & 0.27  & 0.47 &  0.91 & 0.78 &  -0.49 \\ \hline
    0.76 & 0.41  & 0.14 &  0.06 & 0.05 &  -0.50 \\ \hline
\end{tabular}
\caption{Diffusivity, $M_0$ and $i_0$ of Sn using PITT (Figure
\ref{fig:shortTimeCurrentVsTimeFit_FirstLithiation}). The two different fitting
methods give similar values for diffusivity. The slope of the log-log plot being 
close to -0.5 suggests lithiation is diffusion limited.}
\label{table:diffusivityCalibrationSn}
\end{table}
\fi

Based on the Cottrell equation \ref{eq:shortTimeCottrell}, the slope in a
log-log plot of current vs time must be $-0.5$. This is the case in Table
\ref{table:diffusivityCalibrationSn}. The Cottrell solution
assumes that the limiting process is diffusion. Table
\ref{table:diffusivityCalibrationSn} thus suggests that for the Sn phase
(Table \ref{table:diffusivityCalibrationSn}), diffusion is rate-limiting.

\subsection{$M_0, i_0$ for Li$_2$Sn$_5$ and Interface mobility $\chi$}

\begin{figure}[H]
\centering
\includegraphics[scale=0.5]{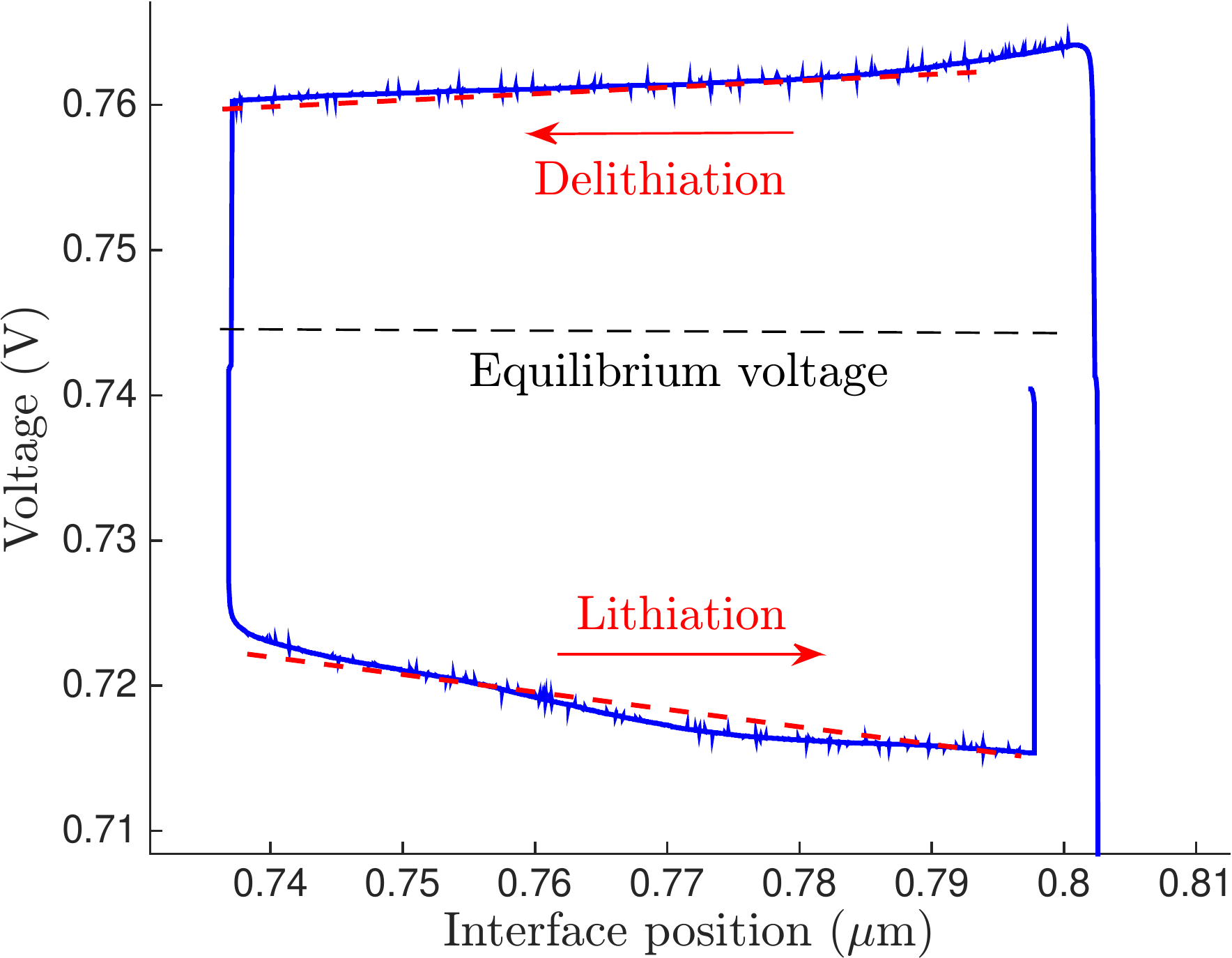}
\caption{Evolution of voltage with interface position during galvanostatic
lithiation/delithiation starting from an interface at equilibrium (blue). The
slope of the line (the linear approximation is shown red) is used to determine
$M_0$ using Equation \ref{eq:galvanostaticM0Calibration}.}
\label{fig:voltageInterfacePositionM0CalibrationCell94}
\end{figure}

To determine $M_0, i_0$ for Li$_2$Sn$_5$ and $\chi$ for the Sn-Li$_2$Sn$_5$
interface, we use galvanostatic and potentiostatic lithiation experiments.
During galvanostatic lithiation, from Equation
\ref{eq:voltageEvolutionGalvanostaticLithiationNondim},
\begin{equation}\label{eq:galvanostaticM0Calibration}
\frac{d\bar{V}}{d\bar{x}_{\text{int}}} = \frac{\bar{I}}{\bar{M}_0}.
\end{equation}
We performed an experiment in which we nucleated the Li$_2$Sn$_5$ phase and
grew it to approximately half the film thickness, let it reach equilibrium, and
lithiated/delithiated at constant current.  Figure
\ref{fig:voltageInterfacePositionM0CalibrationCell94} shows the voltage
evolution as a function of the interface position. The interface position is
calculated from the total charge assuming the Li concentrations in the Sn and
Li$_2$Sn$_5$ phases are equal to their equilibrium concentrations. From the
slope of the linear fits and Equation \ref{eq:galvanostaticM0Calibration}, $M_0
= 9 \times 10^{-19}$ cm$^2$sec${^-1}$ (J mol$^{-1}$)$^{-1}$.

\begin{figure}[H]
\centering
\begin{subfigure}[t]{0.49\textwidth}
\includegraphics[width=\textwidth]{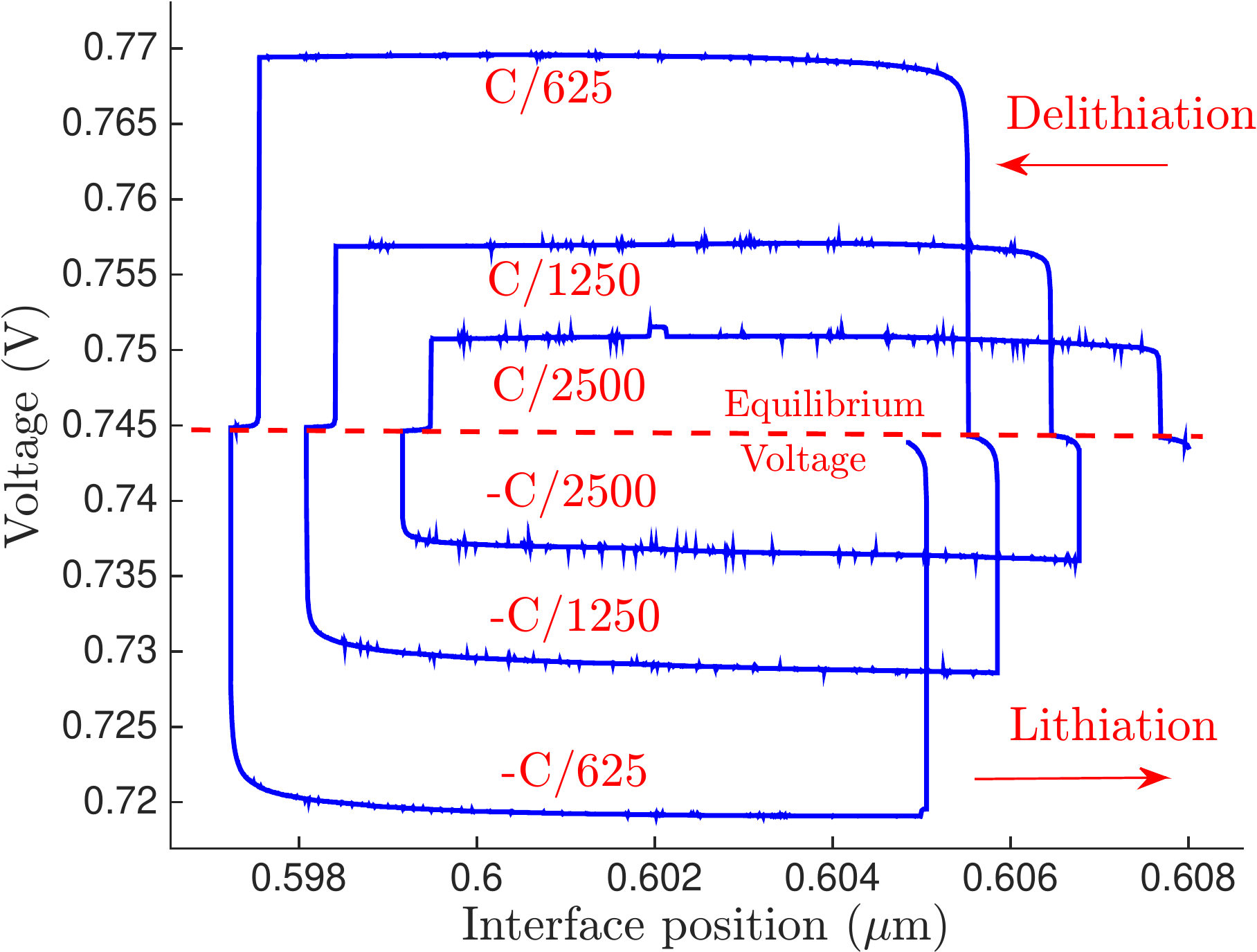}
\subcaption{}
\label{fig:interfacePositionVoltageGalvanostaticCyclingCell94}
\end{subfigure} 
\begin{subfigure}[t]{0.49\textwidth}
\includegraphics[width=\textwidth]{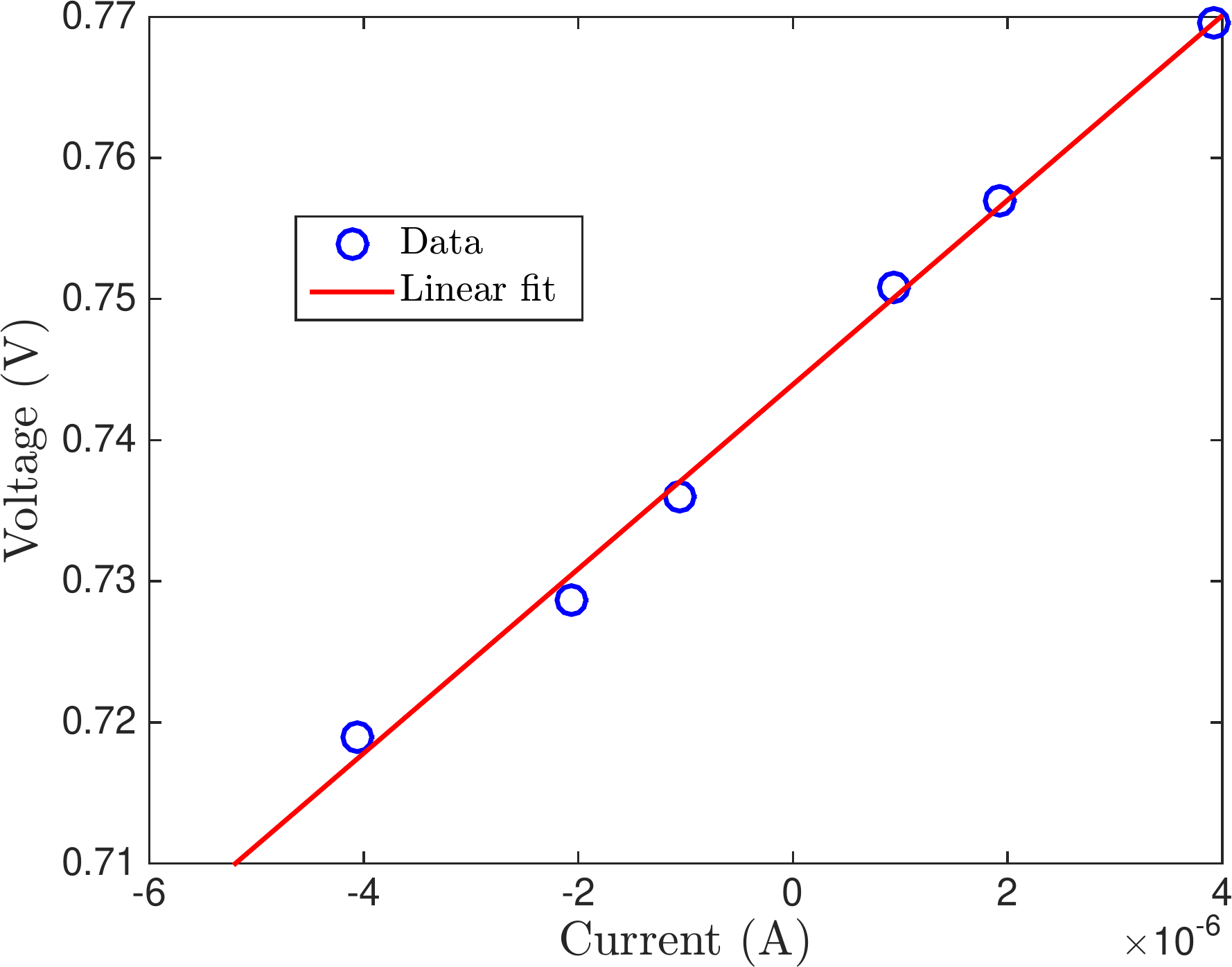}
\subcaption{}
\label{fig:currentVoltageGalvanostaticCyclingCell94_2}
\end{subfigure} 
\caption{(a) Galvanostatic lithiation/delithiation at different currents starting from
an interface at equilibrium. (b)
The plateau potential in (a) varies approximately linearly with the current. The
slope of this line is used to determine the ranges for $i_0,\chi$ (Table
\ref{table:calibrationInterfaceMobilityi0}).}
\label{fig:Cell94GalvanostaticCycling}
\end{figure}

The determine $i_0$ and $\chi$, we repeat the  galvanostatic lithiation
experiments at different currents from C/2500 to C/625 (Figure
\ref{fig:interfacePositionVoltageGalvanostaticCyclingCell94}). The plateau
voltage varies approximately linearly with the current (Figure
\ref{fig:currentVoltageGalvanostaticCyclingCell94_2}). From Equation
(\ref{eq:voltageEvolutionGalvanostaticLithiationNondim}), the slope of this
line is given by,
\begin{equation}\label{eq:slopeVoltageCurrentGalvanostaticNondim}
 \text{Slope } =
\frac{\bar{x}_{\text{int}}}{\bar{M}_0}+\frac{\bar{\chi}}{3\bar{M}_0} +
\frac{1}{\bar{i}_0}.
\end{equation}
The effects of $i_0$ and $\chi$ are indistinguishable in the galvanostatic and
potentiostatic experiments (see Equations
\ref{eq:voltageEvolutionGalvanostaticLithiationNondim} and
\ref{eq:slopeVoltageCurrentGalvanostaticNondim}).  Thus, from the slope and
knowing $M_0$ and $x_{\text{int}}$ (estimated from the total charge), we can
determine the possible ranges for $i_0$ and $\chi$ by setting $\chi = 0$ and
$i_0 \to \infty$ respectively (Table
\ref{table:calibrationInterfaceMobilityi0}).

\begin{table}[H]
\centering
\begin{tabular}{|c|c|c|}
\hline
	 $M_0 $ (cm$^2$sec${^-1}$ (J mol$^{-1}$)$^{-1}$) & Minimum $i_0$ ($\mu$A cm$^{-2}$) & Max $\chi$ ($\mu$m) \\ \hline
    9  $\times 10^{-19}$ & 49.8 & 0.40 \\ \hline
\end{tabular}
\caption{$M_0, i_0,$ for Li$_2$Sn$_5$ and $\chi$ determined from galvanostatic
experiments (Figure \ref{fig:Cell94GalvanostaticCycling}).}
\label{table:calibrationInterfaceMobilityi0}
\end{table}

\begin{figure}[H]
\centering
\begin{subfigure}[t]{0.49\textwidth}
\includegraphics[width=\textwidth]{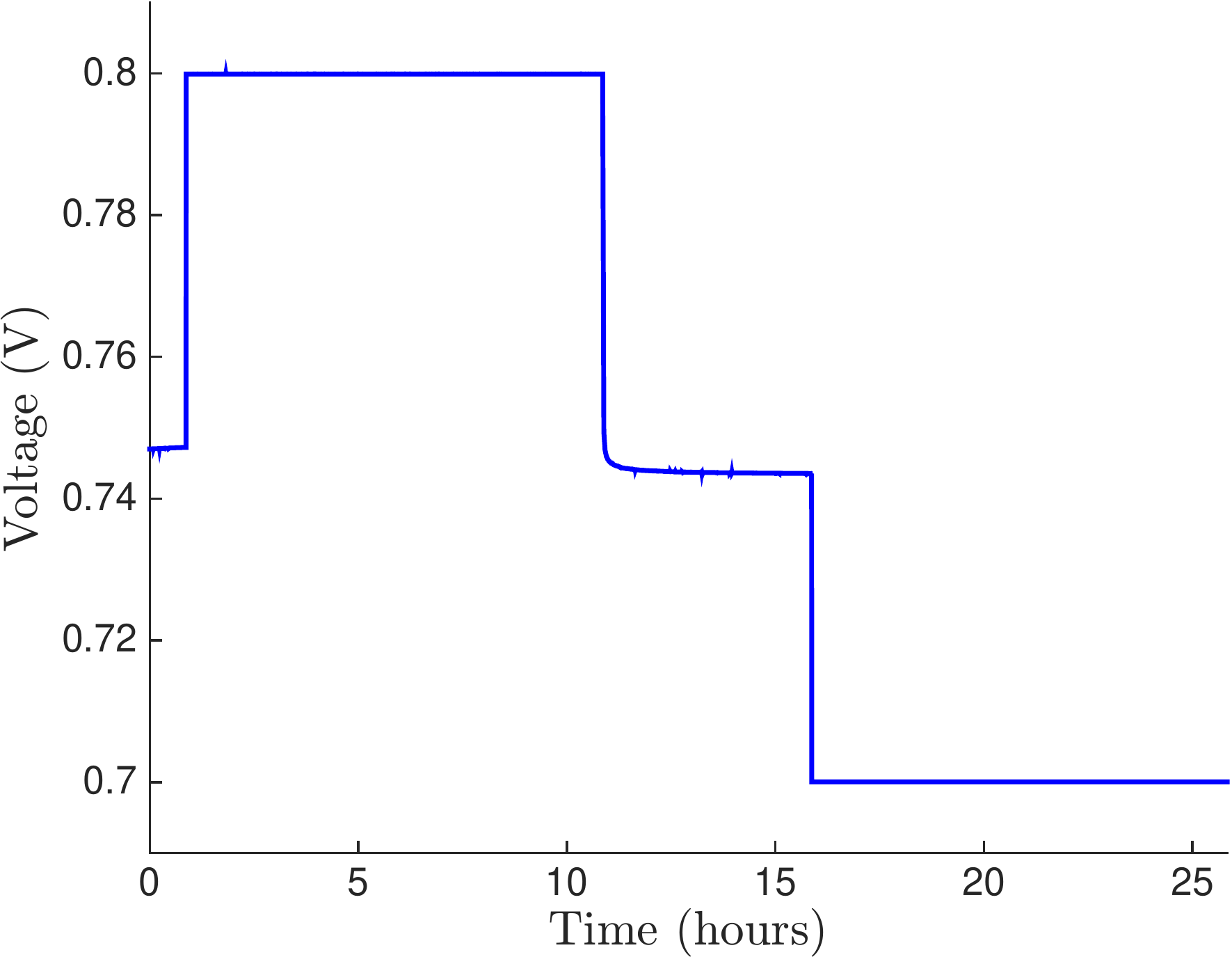}
\subcaption{}
\label{fig:voltageTimePotentiostaticExperimentCahnHilliardComparisonCell94}
\end{subfigure} 
\begin{subfigure}[t]{0.49\textwidth}
\includegraphics[width=\textwidth]{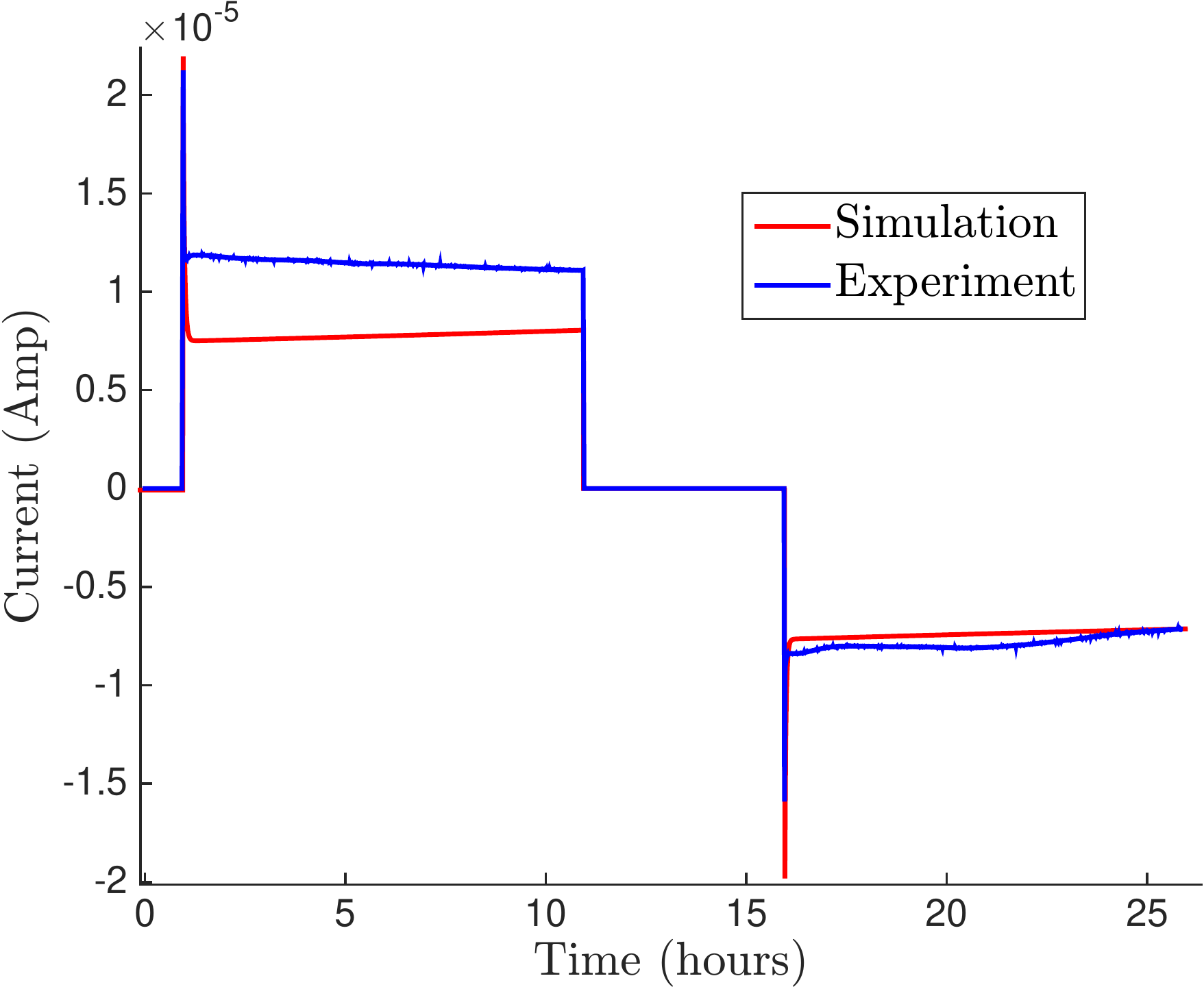}
\subcaption{}
\label{fig:currentTimePotentiostaticExperimentCahnHilliardComparisonCell94}
\end{subfigure} 
\caption{Starting with an interface at equilibrium, voltage steps are applied
to lithiate/delithiate the film. (a) Applied voltage vs time and (b)
corresponding current evolution in experiment (blue) and simulation (red). The
peak current following a voltage step from equilibrium depends only on $i_0$.
The peak current following the first voltage step is used to determine $i_0$ to
be 60 $\mu$A cm$^{-2}$.}
\label{fig:potentiostaticExperimentCahnHilliardComparisonCell94}
\end{figure}

To determine $i_0$, we compare experiments and simulations of potentiostatic
lithiation starting from an interface at equilibrium. The peak current
following a voltage jump depends only on $i_0$ (Figure
\ref{fig:potentiostaticExperimentCahnHilliardComparisonCell94}). From this, we
get $i_0 \approx 60 \mu$A cm$^{-2}$. Using this $i_0$, we get $\chi$ to be 0.07
$\mu$m (using Equation (\ref{eq:slopeVoltageCurrentGalvanostaticNondim})).

\subsection{Simulation with calibrated parameters}

\begin{table}[H]
\centering
\begin{tabular}{|c|c|}
\hline
\textbf{Parameter}  & \textbf{Value}  \\ \hline
Free-energy parameters & \specialcell{$c_0^\alpha = 0.0346, c_0^\beta = 0.4853,
\mu^\text{eq} = 0.193 \text{MJ mol}^{-1}, W = 4.57 \text{MJ mol}^{-1}.$ \\ (Equation
\ref{eq:doubleWellFreeEnergy})} \\ \hline
$M_0$ for Sn                         &   \specialcell{0.14--0.76 $\times
10^{-21}$cm$^2$sec${^-1}$ (J mol$^{-1}$)$^{-1}$ \\ (Equation
\ref{eq:concentrationGradientKineticParameter})} \\ \hline
$D$ for Sn                           &   0.05--1.56 $\times 10^{-16}$ cm$^2$sec${^-1}$ \\ \hline
$i_0$ for Sn                         &   \specialcell{0.29--0.41 $\mu$A cm$^{-2}$ \\ 
(Equation \ref{eq:butlerVolmer})}\\ \hline 
$M_0$ for Li$_2$Sn$_5$               &   9 $\times 10^{-19}$cm$^2$sec${^-1}$ (J mol$^{-1}$)$^{-1}$\\ \hline
$D$ for Li$_2$Sn$_5$                 &   4 $\times 10^{-12}$ cm$^2$sec${^-1}$ \\ \hline
$i_0$ for Li$_2$Sn$_5$               &  60 $\mu$A cm$^{-2}$ \\ \hline
$\chi$ for Sn-Li$_2$Sn$_5$ interface &  \specialcell{0.07 $\mu$m\\
(Equation \ref{eq:concentrationGradientKineticParameter})}\\ \hline
\end{tabular}
\caption{Parameters in the Cahn-Hilliard model calibrated using experiments for
the Sn $\leftrightarrow$ Li$_2$Sn$_5$ transformations.}
\label{table:calibratedParameters}
\end{table}

Table \ref{table:calibratedParameters} shows all the parameters in the
Cahn-Hilliard model calibrated using experiments for
Sn$\leftrightarrow$Li$_2$Sn$_5$ transformations. Using these parameters, we
compare simulations of the Cahn-Hilliard equations with experiments.

\begin{figure}[H]
\centering
\begin{subfigure}[t]{0.49\textwidth}
\includegraphics[width=\textwidth]{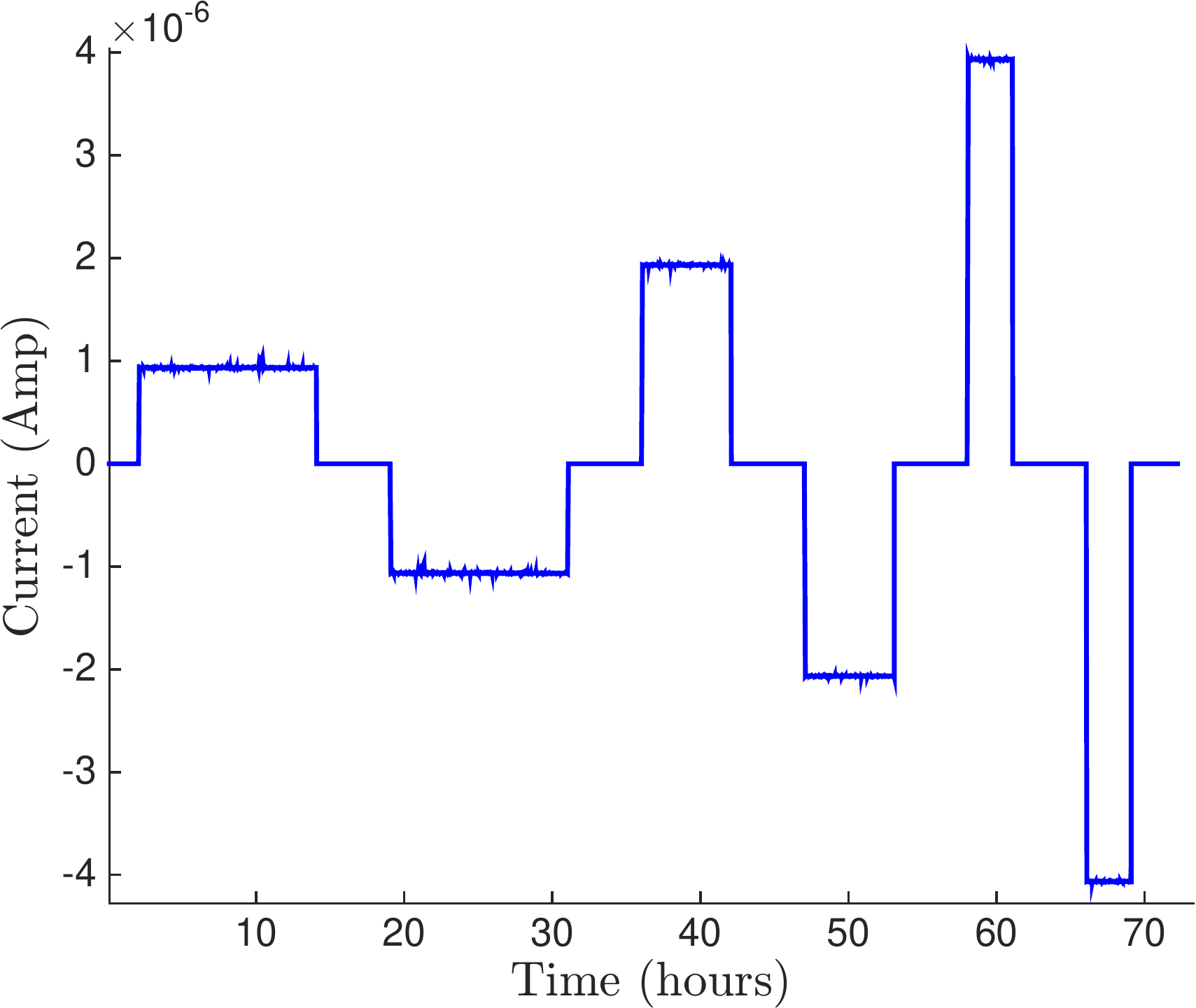}
\subcaption{}.
\label{fig:galvanostaticExperimentCahnHilliardComparisonCurrentvsTimeCell94}
\end{subfigure} 
\begin{subfigure}[t]{0.49\textwidth}
\includegraphics[width=\textwidth]{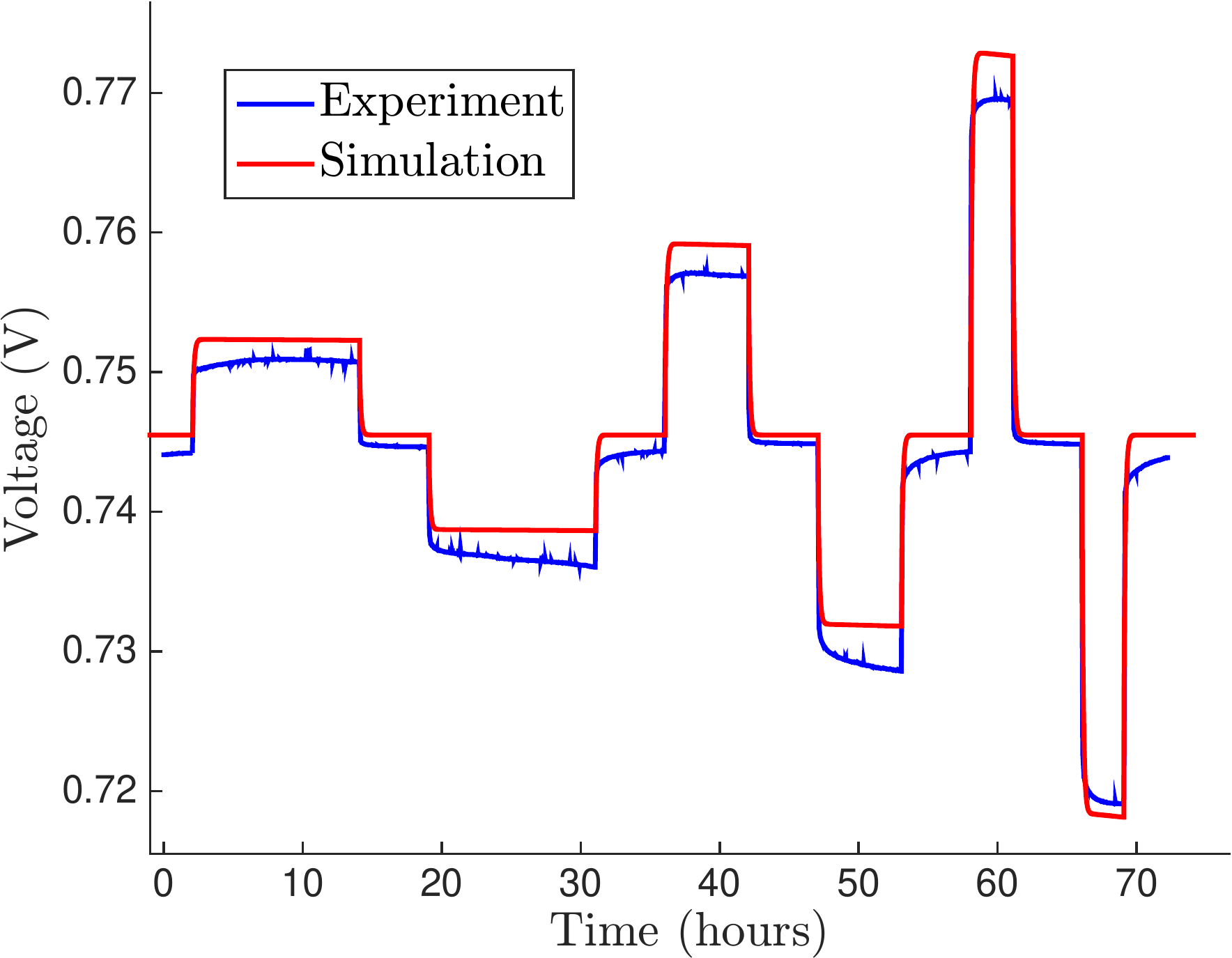}
\subcaption{}.
\label{fig:galvanostaticExperimentCahnHilliardComparisonVoltagevsTimeCell94}
\end{subfigure} 
\caption{Starting with an interface at equilibrium, we lithiate and delithiate
galvanostatically at different currents allowing the film to go back to
equilibrium between successive steps. (a) Applied current vs time and (b) the
resulting voltage evolution in experiment and simulation.}
\label{fig:galvanostaticComparisonCell94}
\end{figure}

Starting with an interface at equilibrium, we lithiate and delithiate
galvanostatically at different currents allowing the film to go back to
equilibrium between successive steps. The applied current is shown in Figure
\ref{fig:galvanostaticExperimentCahnHilliardComparisonCurrentvsTimeCell94} and
the resulting voltage, in both experiment and simulation, is shown in Figure
\ref{fig:galvanostaticExperimentCahnHilliardComparisonVoltagevsTimeCell94}.
The deviations of the voltage from equilibrium in the simulations show good
agreement with experiments. 

\begin{figure}[H]
\centering
\begin{subfigure}[t]{0.49\textwidth}
\includegraphics[width=\textwidth]{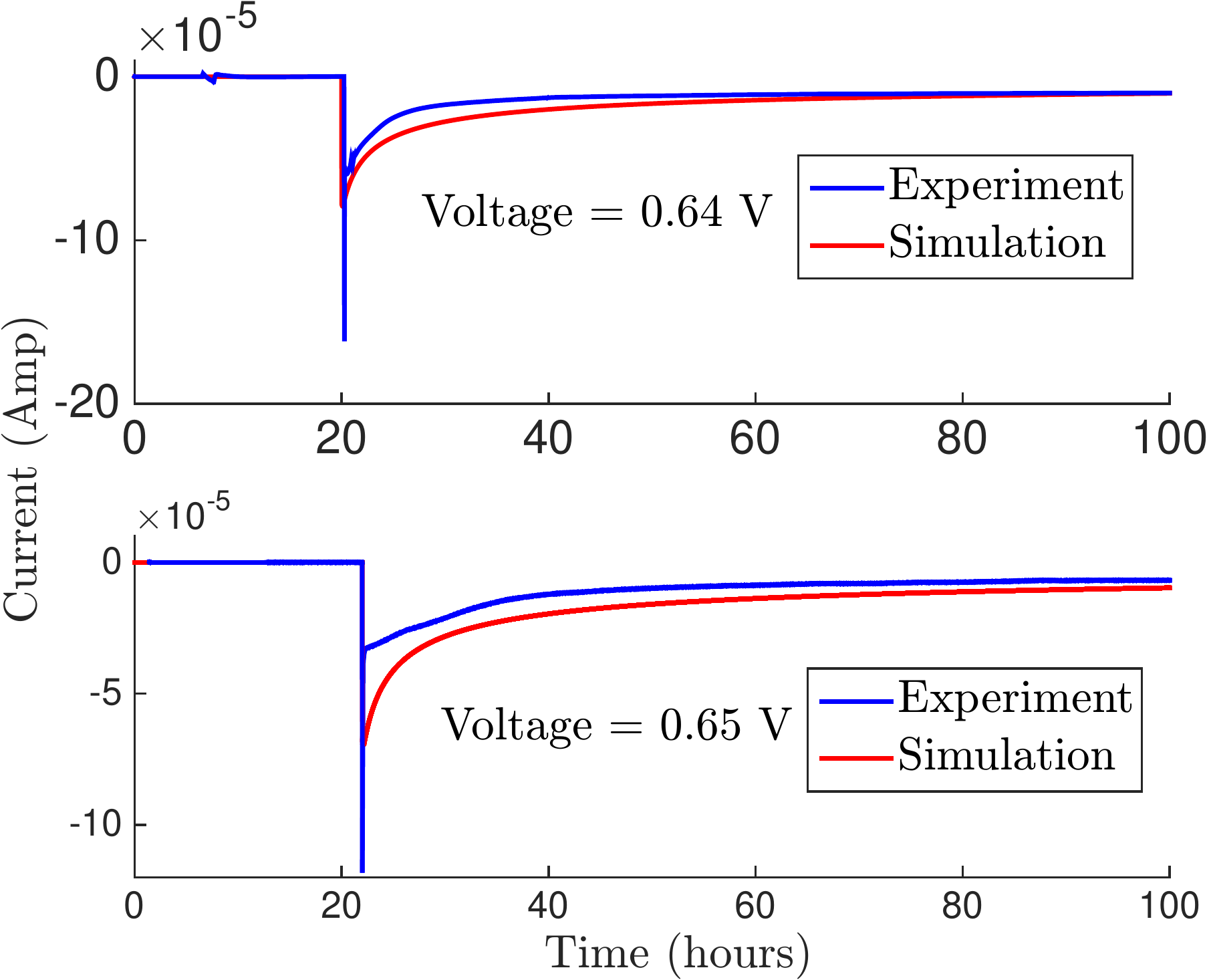}
\subcaption{}.
\label{fig:potentiostaticExperimentCahnHilliardComparisonCells101103}
\end{subfigure} 
\begin{subfigure}[t]{0.49\textwidth}
\includegraphics[width=\textwidth]{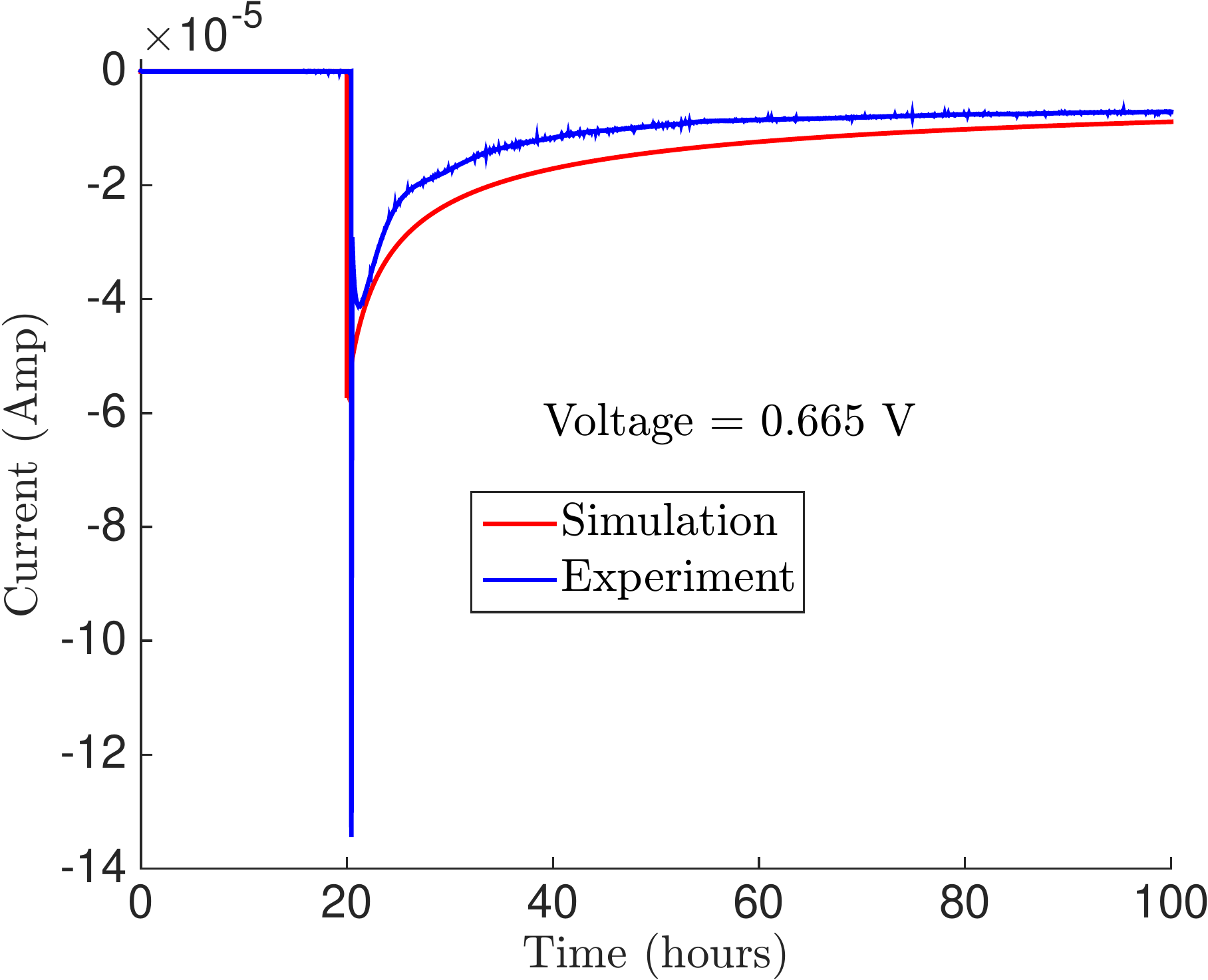}
\subcaption{}.
\label{fig:potentiostaticExperimentCahnHilliardComparisonCell99}
\end{subfigure} 
\caption{Comparison of current evolution in potentiostatic experiments and
simulations three different voltages. The simulations underpredict the peak
current following the voltage step due to the fact that a new phase is not
nucleated in simulations while the subsequent decay closely matches the
experiment.} 
\label{fig:potentiostaticCells99101103}
\end{figure}

Figure \ref{fig:potentiostaticCells99101103} compares potentiostatic lithiation
experiments and simulations of the same at three voltages, 0.64 V, 0.65 V, and
0.665 V. In the experiment, the film was initially in the Sn phase. On lowering
the voltage, the Li$_2$Sn$_5$ phase nucleated and propagated into the film. In
the simulations, we started with two existing phases with the interface close
to the boundary (since the nucleation potential based on the double-well
free-energy is about twice that in experiment. See Figure
\ref{fig:homogenousFreeEnergyExperimentsAfterSEICorrection}). The simulations
underpredict the peak current following the voltage step due to the fact that a
new phase is not nucleated in simulations. Figure
\ref{fig:potentiostaticExperimentCahnHilliardComparisonCell99} shows one of the
experiments (0.665 V). The subsequent decay of the current in the model closely
matches the experiment.

\iftrue
\section{Discussion}\label{sec:discussion}

In this paper, we have calibrated a modified Cahn-Hilliard model for Li-Sn
electrodes that experience a transformation from Sn to Li$_2$Sn$_5$. Although
Cahn-Hilliard models have been used widely in studying battery electrodes and
have been very helpful in understanding qualitative features of microstructure
evolution, not many attempts have been made to make quantitative predictions.
For example, our calibration of the quartic double-well free-energy (Equation
\ref{eq:doubleWellFreeEnergy}) shows that it is impossible to predict correctly
the stoichiometric concentration of the Sn and Li$_2$Sn$_5$ phases, the plateau
potential, and the nucleation potential for the Sn$\rightarrow$Li$_2$Sn$_5$
transformation (Section \ref{sec:calibration}). We expect the other commonly
used free-energy model with the logarithmic free-energy of mixing to have
similar shortcomings. This calls for a more flexible way of including
experimentally determined free-energy into Cahn-Hilliard simulations.

We highlight the significance of interface mobility with the example of
Sn-Li$_2$Sn$_5$ and show that standard Cahn-Hilliard equations result in
infinite interface mobility. One way of including a finite interface mobility
within the Cahn-Hilliard framework is to use a concentration-gradient dependent
Li mobility $M$. The numerical solution of the modified Cahn-Hilliard equations requires
only minor changes in codes for the standard Cahn-Hilliard model (see Appendix). The
perturbation analysis (Section \ref{subsec:perturbationAnalysis}) is very
insightful in understanding the nonequilibrium interface behavior and in
calibration of interface constants.

The connection we made in Section \ref{sec:sharpInterface} between a general
sharp-interface model and the sharp limit of our modified Cahn-Hilliard model
should be a guide in deriving more general Cahn-Hilliard models. The two
nonequilibrium processes at the interface characterized by the constants $K_1$
and $K_2$ (Equation \ref{eq:kineticRelations}) correspond to different
processes. $K_1$ relates to the flux of Li through an interface at rest while
$K_2$ is related to interface motion, transforming material ahead of the
interface from one phase to another.  In the modified Cahn-Hilliard model
presented here, the two are related as $K_2 = 3 K_1$. In general, we do not
expect the two to be related in this way.

In the Cahn-Hilliard modeling of Li-ion battery electrode materials, not much
attention has been paid to the role of interface behavior. The modified Cahn-Hilliard and
its generalizations along with careful experiments should be useful in a wide
range of scenarios where interface mobility is important, both within and
beyond the study of battery electrodes.

\subsection{Implications for battery design}\label{subsec:implications}

\subsubsection*{Rate-limiting process}

The three timescales in the modified Cahn-Hilliard model corresponding to bulk diffusion,
the interface response, and the surface-reaction are given by: 
\begin{equation}
\tau_{\text{diff}} = \frac{H^2}{M_0 W}, \quad \tau_{\text{int}} =
\frac{\chi^2}{M_0 W}, \quad \tau_{\text{surf}} \propto
\frac{1}{i_0}.
\end{equation} 
Depending on the values of the parameters for a
particular system, we can determine the rate-limiting step during the
charge/discharge process. Equating the diffusion and interface
timescales, we get 
\begin{equation}
\tau_{\text{diff}} =\tau_{\text{int}} \implies H = \chi.
\end{equation}

Thus, for films around and smaller than the size $\chi$, the interface behavior
plays an important role in determining the lithiation/delithiation response.
Thus, in the design of electrodes of these sizes, one must give interface
mobility a careful consideration. Finite interface mobility is likely to be
important in other electrode materials that undergo phase transformations.  

\subsubsection*{Interfacial dissipation}

Apart from the dynamics of the charge/discharge process, interfacial mobility
can be important in deciding the energy efficiency of the battery.  Interface
propagation is a dissipative process and the energy lost subtracts from the
useful energy stored in the battery. For simplicity, let us consider the case
when the electrode is being charged/discharged at a constant flux $J$.
Further, assume that the current is small enough that diffusion is at
steady state. In this case, the energy dissipation rates due to diffusion and
interface propagation are given by Equation (\ref{eq:bulkDissipation}) with $j
= J$ for $x$ in $0$ to $a$ and $0$ otherwise and Equation
(\ref{eq:interfacialDissipation}) with $j^- = J, j^+ = 0$. Using these,
\begin{equation}\dot{U}_{\text{int}} = -\frac{AJ^2}{2} \left(\frac{1}{K_1} +
\frac{1}{K_2} \right), \dot{U}_{\text{bulk}} =
-\frac{AJ^2a}{M_0}.
\end{equation}
In charging/discharging the electrode completely, the interface moves between
$x=0$ and $x=H$. The velocity of the interface is given by $v = J/\Delta c_0$.
Integrating the above dissipation rate, the total energy dissipated is given by
\begin{equation}
U_{\text{int}} = \frac{AJH\Delta c_0}{2} \left(\frac{1}{K_1} +
\frac{1}{K_2} \right),  U_{\text{bulk}} =  \frac{AJH^2 \Delta
c_0}{2M_0}.
\end{equation}
Equating the two, the interfacial dissipation is as significant as the bulk one
when 
\begin{equation}H = M_0\left(\frac{1}{K_1} + \frac{1}{K_2}
\right).
\end{equation}
For the modified Cahn-Hilliard case, this becomes $H = \chi/3$ (see Equation
\ref{eq:cahnHilliardKineticParameters}). For systems with characteristic length
scale larger than this, bulk dissipation is important while for smaller
systems, the interfacial dissipation is dominant.
\fi

\iftrue
\section{Conclusions}

In this paper, we used experiments, numerical simulations, and analytical
calculations to calibrate a modified Cahn-Hilliard model for Li-Sn thin film
electrodes. PITT, Potentiostatic, and Galvanostatic experiments were conducted
on Sn thin films measuring transient current and voltage evolution. A modified
one-dimensional Cahn-Hilliard equation along with the Butler-Volmer equation
for the insertion reaction was used to model the experiments. Comparing model
predictions and experiments, we determined the equilibrium and kinetic
properties of the Sn and Li$_2$Sn$_5$ phases and their phase boundary.  The
main conclusions of this study are: 

\begin{itemize}
\item The standard Cahn-Hilliard model captures the nucleation of phases and
diffusion of Li but results in infinite mobility of the phase boundary (Section
\ref{sec:modelBehavior}).
\item A concentration-gradient dependent kinetic parameter can be used to give
the interface a finite mobility and model interface-limited processes.
Perturbation analysis of Cahn-Hilliard equations reveals that the kinetic relations
implied by them are a particular special case of a more general
class. This is useful in developing more general Cahn-Hilliard models that properly
capture the interface behavior (Sections \ref{sec:sharpInterface} and
\ref{sec:modelBehavior}).
\item Analytical double-well free-energies such as Equation
\ref{eq:doubleWellFreeEnergy} though helpful for analysis and understanding
qualitative behavior, can be significantly different from that determined by
experiments (Figure
\ref{fig:homogenousFreeEnergyExperimentsAfterSEICorrection}). Predictive
phase-field models require a better way of incorporating free-energy determined
from experiments.
\item $M_0$ of Sn and Li$_2$Sn$_5$ differ by about 3 orders of magnitude which
suggests that $M_0$ is concentration dependent.
The diffusivity of Li in Sn is around 10$^{-16}$ cm$^2$sec${^-1}$ and in
Li$_2$Sn$_5$ is around 10$^{-12}$ cm$^2$sec${^-1}$ (Tables
\ref{table:diffusivityCalibrationSn} and \ref{table:calibratedParameters}). For
Sn, this is slightly smaller than that reported in \cite{RN175}. We know of no
diffusivity measurements in Li$_2$Sn$_5$.
\item The exchange current density $i_0$ for Sn is of the order $10^{-7}$A
cm$^{-2}$ (Table \ref{table:diffusivityCalibrationSn}) and for Li$_2$Sn$_5$ is
of the order $10^{-5}$A cm$^{-2}$ (Table
\ref{table:calibrationInterfaceMobilityi0}). As far as we know, this is the
first measurement of $i_0$ for Li-Sn. This suggests that for nanometer scale
films, insertion reaction will be rate limiting compared to diffusion (for the
Li$_2$Sn$_5$ phase).
\item The interface-mobility parameter $\chi$ (Equation
\ref{eq:concentrationGradientKineticParameter}) for the Sn-Li$_2$Sn$_5$
interface is 0.07 $\mu$m (Table \ref{table:calibratedParameters}). This is
important since in electrodes at length-scales of $\chi$, the interface
behavior is rate limiting and contributes significantly to the total energy
hysteresis (Section \ref{subsec:implications}). This calls a more accurate
method of measuring the interface mobility and possibly other ways of
characterizing the nonequilibrium interface behavior.
\end{itemize}

\section{Acknowledgement}
This work was supported by the U.S. Department of Energy through DOE EPSCoR
Implementation Grant no. DE-SC0007074.

\begin{appendices}

\section{Finite elements for the modified Cahn-Hilliard equations}

Here we give a brief summary of the finite element method used in solving the
Cahn-Hilliard equations. We can express the governing equations (Equations
\ref{eq:cahnHilliard1d}) in weak form as
\begin{equation} \int_0^H \mu \delta\mu = \int_0^H \frac{dG_0}{dc} \delta\mu + \kappa
\int_0^H \frac{dc}{dx}\frac{d\delta\mu}{dx} \end{equation}

\begin{equation} \int_0^H \frac{\partial c}{\partial t} \delta c = -\int_0^H M
\frac{d\mu}{dx}\frac{d\delta c}{dx} + J \delta c(0).\end{equation}
Some of the boundary terms vanish because of the boundary conditions.
Introducing interpolation functions $\mu = N^a \mu^a$ and $c = \bar{N}^a c^a$, and adopting a 
semi-implicit time integration scheme we obtain the following nonlinear equation system for 
$\mu^a, \Delta c^a$. 

\begin{equation}
\left[ \begin{array}{c} R_a^\mu \\ R_a^c \end{array} \right] = \left[
\begin{array}{c} \int_0^H \left[ \left(\mu +\Delta\mu - \frac{dG_0(c+\Delta
c)}{dc} \right) N^a - \kappa \frac{d(c+\Delta c)}{dx}\frac{dN^a}{dx} \right] \\
\int_0^H \left[ \frac{\Delta c}{\Delta t}\bar{N}^a + \frac{M_0}{1+
\frac{\chi}{\Delta c_0} |\frac{dc}{dx}|} \frac{d(\mu+\theta\Delta
\mu)}{dx}\frac{d\bar{N}^a}{dx}\right] \end{array} \right] + \left[
\begin{array}{c} 0 \\ -J\bar{N}^a(0) \end{array} \right]
\end{equation}
where $0 < \theta < 1$. We have used $\theta = 0.75$ in all our simulations.
The free-energy term is nonlinear, so these equations must be solved using
Newton-Raphson iteration.  The linear equations for the corrections $d\mu^b,dc^b$ 
have the form

\begin{equation}
\begin{bmatrix} K^{\mu\mu}_{ab} & K^{\mu c}_{ab} \\ K^{c\mu}_{ab} & K^{cc}_{ab}
\end{bmatrix} \left[ \begin{array}{c} d\mu^b \\ dc^c \end{array} \right] =
-\left[ \begin{array}{c} R^\mu_a \\ R^c_a \end{array} \right]
\end{equation}
The element stiffness and residual can be expressed in compact form by writing

\begin{equation}
\left[ \begin{array}{c} \mu \\ d\mu/dx \\ c \\ dc/dx \end{array} \right] =
[B][\phi]
\end{equation}
where $[\phi]$ is the nodal degree of freedom vector and $[B]$ is the usual
element interpolation matrix.  The stiffness and residual can then be expressed
as matrix operations
\begin{equation} [K] = [B]^T[D][B], \quad [R]=[B]^T[P] \end{equation}
where 
\begin{equation} [D] = \begin{bmatrix} 1 & 0 & -d^2G_0/dc^2 & 0  \\ 
0 & 0 & 0 & -\kappa \\
0 & 0 & 1/\Delta t & 0 \\
0 & \theta M & 0 & 0 \end{bmatrix} \quad  
[P] = \begin{bmatrix}  \mu +\Delta\mu - dG_0(c+\Delta c)/dc  \\ 
-\kappa d(c+\Delta c)/dx \\
\Delta c/\Delta t \\
Md(\mu+\theta\Delta \mu)/dx \end{bmatrix} \end{equation}
Combining element degrees of freedom in the usual way yields a standard
nonlinear finite element system of equations
\begin{equation} R(u) = F\end{equation}
which are solved by Newton-Raphson iteration. The correction $\Delta u^k$ to
the degree of freedom vector $u$ at the $k$th iteration is obtained by solving
the linear system
\begin{equation} K \Delta u^k = F - R(u^{-1}) \end{equation}
where $K$ and $R$ are determined by assembling the element stiffness and
residuals defined above.

The Butler-Volmer equation (Equation \ref{eq:butlerVolmer}) can be included
within the standard finite element framework by adding an additional node to
the mesh, which has the voltage V as its degree of freedom; and adding an
element to the mesh which connects the voltage node and the node at the surface
of the mesh.  The element has a generalized force vector
\begin{equation} [R] = \begin{bmatrix}  J(V,\mu)  \\ 0 \\ 0 \end{bmatrix}\end{equation}
where $J$ is the flux calculated from the Butler-Volmer equation.
The corresponding stiffness is
\begin{equation} [K]  =  \begin{bmatrix} \partial J/\partial V & \partial J/\partial V &  0  \\ 
0 & 0 & 0 \\ 0 & 0 & 0 \end{bmatrix} \end{equation}

\end{appendices}

\bibliographystyle{unsrt}

\fi

\end{document}